# Recent progress of fabrication, characterization, and applications of anodic aluminum oxide (AAO) membrane: A review

Saher Manzoor[1], M. Waseem Ashraf[1,*], Shahzadi Tayyaba[2], M. Khalid Hossain[3,4,**]

[1]*Dept. of Physics (Electronics), GC University, Lahore 54000, Pakistan*
[2]*Dept. of Computer Engineering, The University of Lahore, Lahore 54000, Pakistan*
[3]*Atomic Energy Research Establishment, Bangladesh Atomic Energy Commission, Dhaka 1349, Bangladesh*
[4]*Interdisciplinary Graduate School of Eng. Sciences, Kyushu University, Fukuoka 816-8580, Japan*
Correspondence: * dr.waseem@gcu.edu.pk; **khalid.baec@gmail.com; khalid@kyudai.jp

## Abstract

The progress of membrane technology with the development of membranes with controlled parameters led to porous membranes. These membranes can be formed using different methods and have numerous applications in science and technology. Anodization of aluminum in this aspect is an electro-synthetic process that changes the surface of the metal through oxidation to deliver an anodic oxide layer. This process results in a self-coordinated, exceptional cluster of round and hollow formed pores with controllable pore widths, periodicity, and thickness. Categorization in barrier type and porous type films, and different methods for the preparation of membranes, have been discussed. After the initial introduction, the paper proceeds with a brief overview of anodizing process. That engages anodic aluminum oxide (AAO) layers to be used as formats in various nanotechnology applications without the necessity for expensive lithographical systems. This review article surveys the current status of the investigation on AAO membranes. A comprehensive analysis is performed on AAO membranes in applications; filtration, sensors, drug delivery, template-assisted growth of various nanostructures. Their multiple usages in nanotechnology have also been discussed to gather nanomaterials and devices or unite them into specific applications, such as nano-electronic gadgets, channel layers, and clinical platforms tissue designing. From this review, the fact that the specified enhancement of properties of AAO can be done by varying geometric parameters of AAO has been highlighted. No review paper focused on a detailed discussion of applications of AAO with prospects and challenges. Also, it is a challenge for the research community to compare results reported in the literature. This paper provides tables for easy comparison of reported applications with membrane parameters. This review paper represents the formation, properties, applications with objective consideration of the prospects and challenges of AAO applications. The prospects may appeal to researchers to promote the development of unique membranes with functionalization and controlled geometric parameters and check the feasibility of the AAO membranes in nanotechnology and devices.

## Graphical Abstract

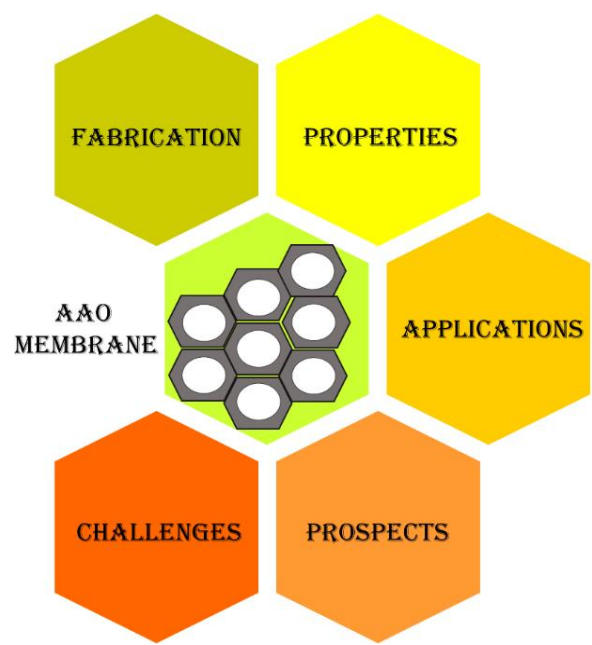

Contents

| List of abbreviations | | | |
|---|---|---|---|
| **AAO** | Anodized Aluminum oxide | **P3HT** | Poly 3-hexylthiophene |
| **EPR** | Electron paramagnetic reverberation | **ECR-CVD** | Electron cyclotron resonance chemical vapor deposition |
| **PB** | Pilling Bedworth | **PPy/DBS** | Polypyrrole doped with the dodecylbenzenesulfonate |
| **SSAW** | Standing surface acoustic waves | **FITC-BSA** | Fluorescein isothiocyanate labelled Bovine serum albumin |
| **PDMS** | Polydimethylsiloxane | **PGMA** | Propylene glycol monoacetate |
| **AT-AAO** | Al textured AAO | **DOX** | Doxorubicin |
| **GZO** | Ga-doped ZnO | **PBS** | Phosphate buffered solution |
| **WG** | Wrinkled graphene | **PDMS** | Polydimethylsiloxane |
| **SERS** | Surface enhanced Raman scattering | **VACNTs** | Vertically aligned carbon nanotube |
| **SMSA** | Simultaneous multi-surfaces anodization | **IAAA** | Indole-3-acetic acid |
| **FPI** | Fabry-Perot interferometer | **LCD** | Liquid crystal display |
| **GNW** | Gold nanowire | **PAN/GO** | Polyacrylonitrile/graphene oxide |
| **TENG** | Triboelectric nanogenerator | **ZIF-8** | Zeolite imidazolate framework 8 |

# 1  Introduction

Since the beginning of the 1940s, membrane technology has advanced as a subdiscipline of physical chemistry and engineering. The development of this technology reached a milestone in the late 40s with the development of microporous membranes. Other landmarks that marched the development of this technology include permionic membranes, asymmetric membranes for reverse osmosis and ultrafiltration, and gas-permeation membranes. Overall, the dyad of research direction and activities for the development of membrane technology can be categorized into two critical stages. The first stage was till the 1980s. Scientists made contributions to the early development of membrane formation using suitable materials. After the 1980s, the second stage is based on membrane formation methods with controlled conditions for improved membrane characteristics [1–5] Several classification schemes have been reported for membranes due to various material, geometry, and fabrication techniques [6]. The placement of membranes can be done into a variety of classes, as shown in **Figure 1**.

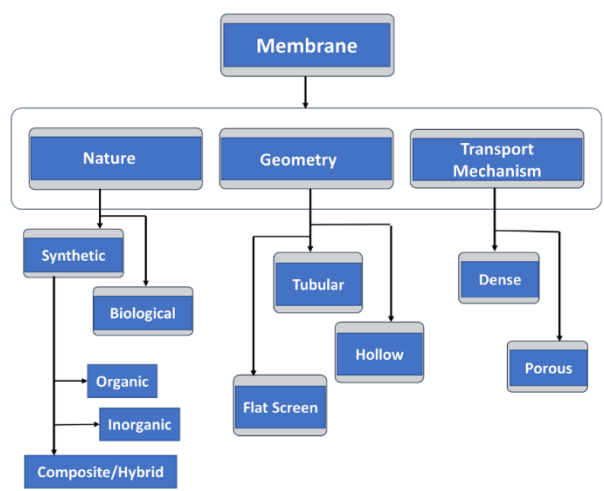

**Figure 1.** Classification of membranes.

Membranes depending upon the features like nature, geometry, and transport mechanism, have numerous medical, and industrial applications like drug delivery [7,8], tissue engineering, dialysis, cold sterilization for pharmaceuticals, water purification, and industrial gas separation [9,10]. Membrane processes have also been proved more valuable than traditional technologies. The reason for the wide range of applications is that the pressure-driven membrane processes provide size-based separation for a wide range from micro to nano [11].

The development process of the nanostructured membrane comprises several challenges and requires the mutual efforts of material scientists and analytical chemists. As for the growth of membrane design and technology, understanding nanostructures, functionalization of nanomaterials, and chemical incorporation play a vital role [10]. Regarding the material of membrane, the researchers performed an extensive investigation on organic membranes in the beginning. But after some time, inorganic membranes gathered their attention. As inorganic membranes show high resistance to temperature and pressure, have a longer life span, and are insensitive to bacterial attack [12,13].

Aluminum was preferred over other materials in many applications due to its outstanding properties. However, most of the applications require well-defined surface properties that fluctuate by different surface treatments. This paper aims to review the aluminum oxide membrane [14–17], its properties, formation process with controllable geometric features, and detailed applications. Various applications of AAO membranes are discussed in detail with the help of available literature, and with each application, promising prospects are given. Finally, the paper concludes with future directions for the development and applications of AAO membranes in various fields. This study may open new paths for investigations in AAO applications.

## 2 Porous alumina membranes production techniques

As far as porous aluminum oxide membrane is concerned, the following commonly used preparation methods are reported in the literature.

### 2.1 Sol-gel method

Aluminum oxide membranes with ultrafine pores can be produced by using the sol-gel method [18–20]. The precursor used for bulk ceramic bodies is boehmite, Alumatrane, or tris(alumatranyloxy-*i*-propyl) amine sol with binders and additives are used during this process for coherency of the body during the process of firing and drying. The basis of this procedure is that sols are used instead of powders to get narrow pore size distribution.

Steps of membrane formation

- Preparation of sol
- Gel formation (drying)
- Calcination
- Sintering [13,21–25]

The schematic diagram for the sol-gel process is shown in **Figure 2** and **Table 1** depicts the influence of prolonging heat treatment during the sol-gel method for aluminum oxide membrane.

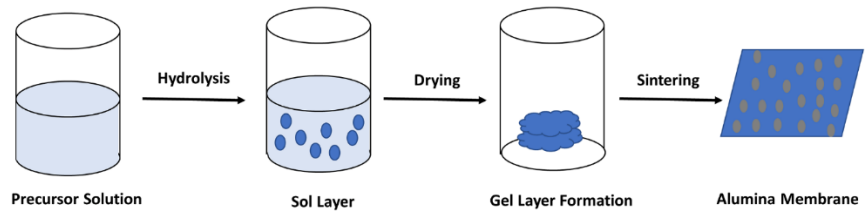

**Figure 2.** Schematic for sol-gel process.

**Table 1** Influence of prolonging heat treatment during sol-gel method for aluminum oxide membrane.

| Sintering Temperature ($^0$C) | Pore size (nm) | Porosity (%) | Phases after thermal treatment |
|---|---|---|---|
| 400 | 4.7 | 53 | γ-Al$_2$O$_3$ |
| 600 | 5.5 | 55 | γ-Al$_2$O$_3$ |
| 700 | 6 | 53 | γ-Al$_2$O$_3$ |
| 900 | 10 | 48 | θ-Al$_2$O$_3$ |
| 1200 | 110 | 41 | α-Al$_2$O$_3$ |

### 2.2 Phase inversion and sintering

In the Phase inversion method for the formation of Aluminum oxide membranes, powder of alumina with specified particle size is used to make precursors in the presence of an organic binder. The precursor is then sintered at high temperatures. The spinning process and sintering temperature are the parameters that affect the pore size and porosity. In contrast, the most crucial factor is the particle size of alumina powder that is to be used.

Steps of membrane formation

- Dope (organic binder) and powdered $Al_2O_3$ are mixed to form the precursor.
- Milling/mixing.
- The suspension is then degassed using a vacuum pump.
- Immersion of the suspension in water for phase immersion
- Sintering at high temperature after drying.

The schematic diagram for the process is given in **Figure 3**.

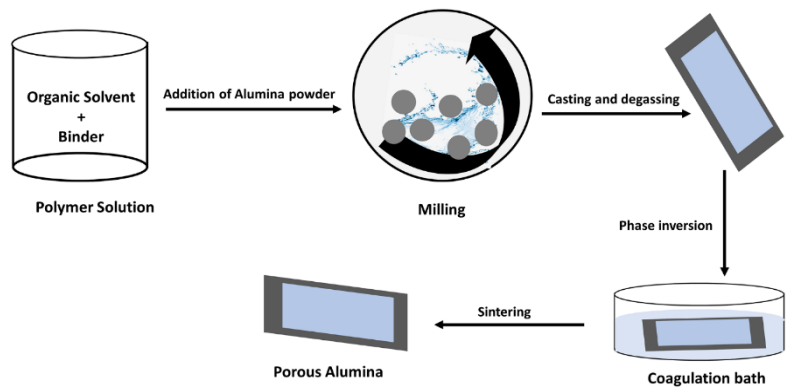

**Figure 3.** Schematic for phase inversion and sintering.

The pores in the produced $Al_2O_3$ membrane are simply voids between the particles, with no defined shape or size. So higher alumina content and smaller particle size lead to the denser membrane and smaller pore size. But need not mention if the quantity is increased beyond a limit, porous membrane cannot be produced using this method [26–30].

## 2.3 Anodization

Aluminum cannot exist in a natural environment in its basic form, and a 1-3 nm thick oxide layer is formed that protects the metal from further reaction with oxygen. Buff [31] in 1857 introduced an electrolytic oxidization process that forms a relatively thick layer than the natural. That enables the use of metal at the industrial level to apply for biomedical degradable devices and sensors [32,33]. In the 1920s, this observed phenomenon of oxidization was exploited for industrial applications [34]. In the 1960s, exposure tests were carried out on AAO in severe atmospheric conditions. The tests revealed that the substrate experiences two types of corrosion, namely pitting and surface bloom. In pitting, small pits appear at the surface. The number of pits depends on the thickness of the film. Thus, by increasing film thickness, the resulting structure can be made corrosion-resistant in almost any environment. While surface bloom is opposite to pitting, it depends on chemical re-attack and degree of sealing. Eq. (1) is presented for the selection of film thickness considering the given factors.

$$U = 2H(P + S) \tag{1}$$

Where U represents the film thickness, H is the humidity grade, P indicates the environmental pollution, and S is the salt content [35].

Weak acids result in anodized films of limited thickness as the films prepared are dark in color and easily break if the voltage exceeds a particular value. Strong anodizing acids produce explicit films at almost constant anodizing voltage. In contrast, intermediate acids work at a stable anodizing voltage. The literature also suggested that we can achieve desired membrane properties by mixing two or three acids to form an electrolyte bath [36]. Further investigation on the matter unveils that the thickness of the oxide layer can change based on the temperature. It is found that at room temperature, thickness is 1-2nm which can grow up to 10-15nm at 500°C. This process is helpful for the protection of metals from abrasion and corrosive chemicals like sulfates like chlorides. This thick layer may keep the metal surface from further oxidation. However, local oxidization of metal may occur due to environmental conditions like changes in pressure and temperature. The process is called anodization, as the electrolytic bath has aluminum as an anode. In the electrolysis process, the Al is soaked into an acid electrolyte bath. Electric current is passed through the bath in the presence of a cathode, as shown in **Figure 4**. The whole reaction can be expressed using the following equation (Eq. (2)):

$$2Al + 3H_2SO_4 \rightarrow Al_2(SO_4)_3 + 3H_2 \qquad (2)$$

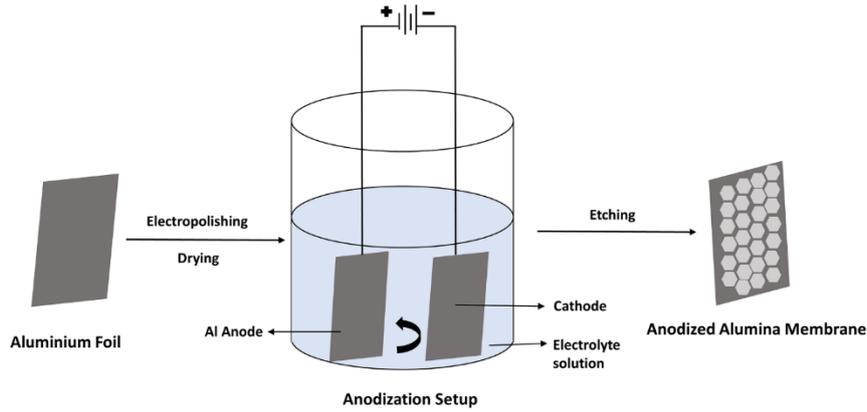

**Figure 4.** Schematic of the anodization process.

Anodization of metals and semiconductors like Al, Ta, Nb, Mg, W, Hf, Zr, Si, Hf, and InP has been reported to form nanoporous structures. The oxidization process produces two types of films depending upon electrolyte and anodizing conditions. These films are known as porous type and barrier type [31,34,37].

### 2.3.1 Porous Films

Porous type thick films are formed when the anodic film is chemically reactive to the electrolyte like phosphoric, sulphuric, or oxalic acid. These types of films have a barrier layer in contact with the metal and a thick porous layer above it. The thickness of the porous layer is directly proportional to the amount of anodizing time.

### 2.3.2 Barrier Films

Barrier-type thin and compact films are formed when the anodic film is insoluble or mildly reactive to the electrolyte like tartrate, phosphate, or neutral borate solutions. The thickness of these films is directly proportional to the voltage applied depending upon the electrolyte type and its concentration. The current density A in the anodic film is given by the high-field conduction equation (Eq. (3)).

$$A = i\exp(TF) \qquad (3)$$

Where i and T are temperature-dependent constants, F is the field strength which V/d can replace, V is the voltage, and d represents the film thickness. These types of films reach a thickness between 1.2-1.44 nmV$^{-1}$ based on different anodizing conditions. Eq. (1) represents the exponential growth of anodic film thickness depending upon the voltage, current density, and film thickness.

The thickness of these films increases with the current density applied; however, anodizing time doesn't affect the thickness between the outer and inner layers. Minimum loss of ions occurs into electrolyte bath from the film while forming barrier-type films. Addition of oxygen and loss of electrons or hydrogen ions. These films are used to make the capacitor, and the aluminum layer protects the capacitor. The properties of barrier and porous oxide films are given in **Table 2** [38–42].

**Table 2** Properties of barrier and porous oxide films [13,21–25].

| Film type | Common Electrolytes | Morphology | Pores Growth | Membrane structure | pH | Applications |
|---|---|---|---|---|---|---|
| Barrier | Phosphate, Tartrate, Ammonium Borate | Amorphous Uniform in thickness Sharply defined | Exponential | Circular | >7 | Capacitors Rectifiers |
| Porous | Sulphuric acid, Oxalic acid, Phosphoric acid | Duplex | Linear w.r.t voltage | Hexagonal cells with cylindrical nanopores | <5 | Paints Lacquers Adhesives |

Both types of layer structures have applications in industry. Barrier type helps make capacitors because of its thin, resistant yet hard layer that behaves as an insulator. On the contrary, Porous films are thick and have a highly controllable density distribution, pore diameter, cylindrical shape, and periodicity. Because of their high aspect ratio and porous structure, these templates are widely used in nanotechnology [38–42].

## 3 Types of anodization process for porous AAO

In the beginning, when anodized aluminum was used in industrial applications, the main focus of the surface finishing industry was to develop a cost-effective anodization process with improved properties of the resulting products. So, the typical approach to get porous AAO membrane used then did not result in an ordered structure free of cracks. Therefore, that process could not be used for applications in nanotechnology. Hence, attempts were made to develop an Anodization process so that the geometric factors can be engineered. AAO membrane is self-coordinated nanopores that have gathered immense attention due to the simplicity of its formation and control over geometric properties. Pore development behavior under various precise conditions has been studied to plan an interesting permeable stage for explicit applications significantly. In this section, all the different types of anodization used to get porous AAO structure are explained briefly.

### 3.1 Hard anodization

The typical hard anodization process adopted by the industries in early 1920 was carried out at a potential more incredible than the breakdown value that resulted in cracks in the resulting anodic structure with poor mechanical stability. Thus, the prepared structure could not be used for practical applications in nanotechnology. It was a challenge to control the parameters of the membrane using that conventional hard anodization process, so various attempts were made to overcome the problem associated with the process. The modified Hard anodization process is carried out at a high voltage, but the voltage value is below the dissolution of the metal. The primary benefit of this type of anodization is that we can get a high-quality nanoporous oxide structure with a fast growth rate [43,44]. Lee et al. applied the hard anodization process for anodic alumina using voltage 110-115V in the presence of oxalic acid as an electrolyte. A perfectly ordered AAO membrane with a high aspect ratio was with an oxide growth rate increase by 70mm/h was reported [36].

### 3.2 Mild anodization

Mild anodization is carried out at relatively low voltage, and thus the oxide growth is slow. In 1996 Masuda et al. developed a two-step anodization process in which the anodization is carried out in two steps. In between these steps, etching is done to remove the distorted portion. The conditions during both steps for anodization are the same except for each step's time interval. The development of this method led to several other studies by researchers for a better understanding of parameters and their dependence on the anodization conditions. The anodization process may be optimized for various applications [45,46].

### 3.3 Pulse anodization

In the pulsed anodization process, low and high potentials are applied in pulses for both hard and mild anodization conditions. Many researchers were combining both hard and mild anodization processes to achieve desired membrane properties. In 1987 G. C. Tu and L Y. Huang reported pulsed AC and DC anodization; however, the results showed that the structure had satisfactory corrosion resistance with cracks and high surface roughness [47]. Lee et al. also used pulses of hard anodization voltage and mild anodization voltage in the presence of sulphuric acid as an electrolyte at room temperature to achieve AAO Membrane [48].

### 3.4 Cyclic anodization

Potentiostatic or galvanostatic periodic oscillatory signal is applied in this type of anodization process. A slow change of voltage/current between the Mild and Hard anodization modes is applied using the periodic signal, resulting in the porous structure. Losic *et al.* applied this type of anodization process for the formation of AAO membrane. Although the concept of cyclic anodization is not new, it has also been used previously for the anodization of sputter-deposited alloys. The parameters chosen for the cyclic anodization result in the variation of the morphology of the anodized membrane [49,50]. The variation in the potential changes during hard and mild anodization for pulse anodization and cyclic anodization is shown in **Figure 5**.

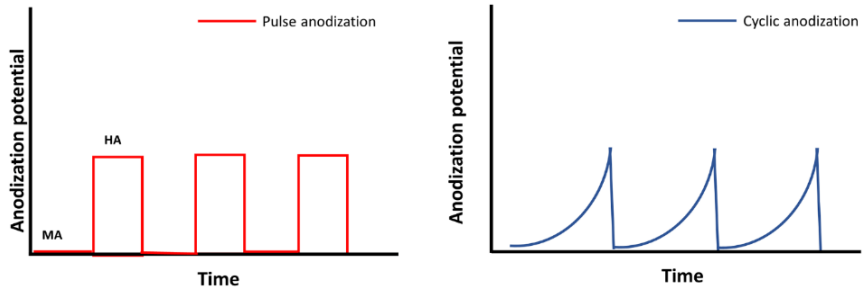

**Figure 5.** Variation of anodization potential versus time. Pulse anodization (left), Cyclic anodization (right).

## 4    Characterization of anodized aluminum oxide

The anodization process results in a nanostructure that consists of a periodically arranged hexagonal pore channel. Nucleation of pores occurs on the substrate and grows in the vertical direction maintaining their direction coherent. The geometric feature of the prepared structure strongly depends on the experimental conditions. Distinct quantities for the characterization of the AAO membrane are pore diameter, pore density, porosity, pore length, and interpore distance. **Figure 6** shows the horizontal view of the AAO membrane with hexagonal cells with pores in their centers, and each hexagonal cell is separated by cell boundary. **Figure 7** represents the 3D dimensional view of AAO membrane with well-defined oxidized hexagonal cells on Aluminum substrate.

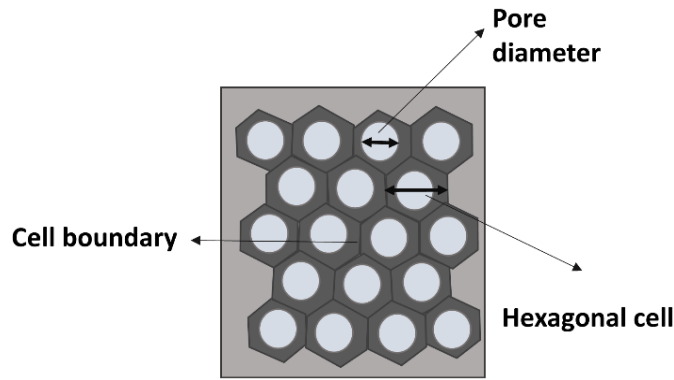

**Figure 6.** Horizontal view of ideal AAO membrane.

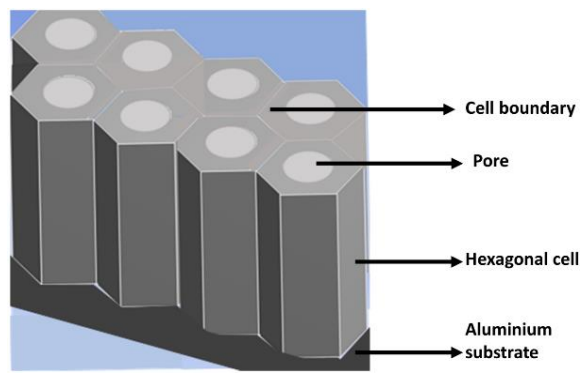

**Figure 7.** 3-dimensional view of AAO.

This section explains the parameters for highly ordered nanostructured AAO membrane characterization and how experimental conditions affect them. The parameters of the membrane depend on the anodization conditions. Other factors that affect the pore formation during anodization are listed.

- The volume of oxide is more prominent than the metal devoured, and subsequently, there are solid strains in the oxide layer that cause mechanical pressure, instating the formation of pores.
- The AAO forming could also suffer from local differences within the wettability of the surface of the oxide.
- The AAO framing could likewise experience the ill effects of nearby contrasts inside the wettability of the oxide surface.

- A part of the Al ions, $Al^{3+}$, is delivered from the metal into solution without binding to the oxide structure. This phenomenon diminishes the current proficiency of AAO response.

- The presence of unfamiliar components in the Al substrate can generate oxygen gas as a side response. Subsequently, it lessens the current effectiveness of the oxidation reaction.

- The thickness of the barrier layer is proportional to the applied voltage, with other parameters being fixed.

The most critical parameters for the characterization of the anodized membrane, i.e., pore diameter, interpore density, porosity, and pore density, are explained.

### 4.1 Pore diameter

The hexagonal cell is the region that constitutes nanopore within, as shown in **Figures 6** and **7**. The pores in the membranes are hollow circular regions whose diameter depends on the anodization conditions. By varying the anodization parameters, the diameter of the pores can be varied. When anodization voltage is increased, the pore diameter also increases as Pilling Bedworth (PB) ratio suggests that pore formation is accompanied by volume expansion (Eq. (4)).

$$\text{PB ratio} = \frac{\text{Volume of formed oxide}}{\text{Volume of metal consumed during oxidation}} \quad (4)$$

When voltage is applied, the pore initiation occurs. With an increase of voltage, more pores initiate on the substrate, and the pore diameter increases resulting in a pear drop shape at high voltage. Similarly, the anodization time has a direct relationship with the pore diameter. On the contrary, if the concentration of the electrolyte increases, fast oxidation occurs, resulting in a decrease in pore size [43–45].

Metal dissolution varies for different voltage types of pore growth during different types of anodization processes is given in **Figure 8**. In hard anodization **Figure 8(a)**, high voltage is applied, so the pores are comparatively thin from the top, whereas in mild anodization **Figure 8(b)**, the upper portion of pores is quite wide. In cyclic and pulse anodization **Figure 8(c)**, MA and HA processes are applied alternatively, so the pore length varies accordingly.

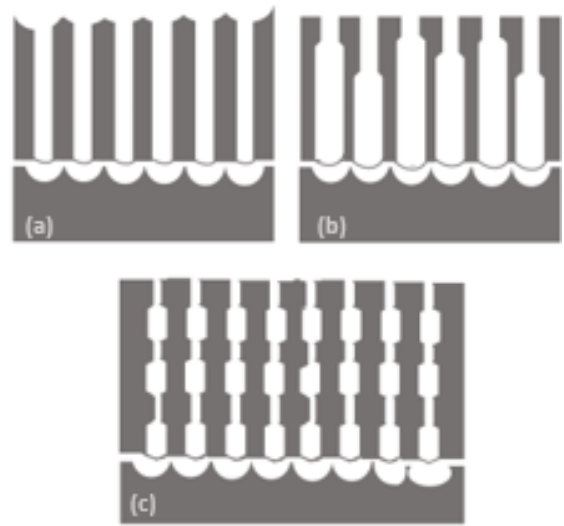

**Figure 8.** Pore growth for different anodization processes (a) hard anodization, (b) mild anodization, (c) cyclic and pulse anodization.

### 4.2 Interpore distance

Distance between the centroids of neighboring pores is called interpore distance $D_{int}$. Eq. 5 shows it is directly proportional to the applied voltage during the anodization process.

$$D_{int} = \delta V \quad (5)$$

Where δ is the proportionality constant with a value of 2.5nm/V.

The relation between pore diameter and the interpore distance is given in Eq. 6:

$$D_{int} = P_D + 2W_t \tag{6}$$

Where $P_D$ is pore diameter, and $W_t$ is the thickness of the pore wall. The average interpore distance remains constant with the increase in electrolyte concentration, and as far as temperature is concerned, it is reported that interpore distance is independent of temperature [41,51–53].

### 4.3 Porosity

The porosity of the membrane can be tailored if the pore diameter and the interpore distance can be varied using specified anodization conditions. The porosity is related to the ratio of pore diameter and interpore distance and can be calculated using Eq. 7.

$$\text{Porosity} = \frac{\pi}{2\sqrt{3}} \frac{\text{Pore diameter}}{\text{Interpore distance}} \tag{7}$$

### 4.4 Pore density

Pore density is related to the number of pores generated by the anodization process and pore diameter. The formula used for the calculation of pre-density is given by Eq. 8.

$$\text{Number of pores} = \frac{A}{\pi/4} (P_D^2) \tag{8}$$

Where A is the area of pores, and **$P_D$ is** the pore diameter. If the pore diameter can be controlled with varying the anodization conditions, then pore density can also be manipulated [51–54].

**Table 3** shows the AAO membrane fabricated using different anodization types and the resulting parameters of the membrane [15,43,45–47,50,54–65].

**Table 3** Effect of anodization type and three mostly used acidic electrolytes on characterization parameters.

| Electrolyte used | Anodization type | Temperature (ºC) | Voltage (V) | Pore size (nm) | Interpore distance (nm) | Substrate thickness (μm) | Refs. |
|---|---|---|---|---|---|---|---|
| **Oxalic acid** | Hard anodization | 1 | 100-150 | 49-59 | 200-300 | 110 | [43] |
| | Hard anodization | - | 130 | 40 | >400 | 500 | [45] |
| | Hard anodization | 0-5 | 140-180 and 140-200 | ~100 | - | Disk of diameter 20000 | [63] |
| | Two-step anodization | 17 | 40 | 50 | 100 | - | [46] |
| | Two-step anodization | 5-7 | 40 and 20-50 | ~70 | - | 500 | [64] |
| | Cyclic Anodization | <0.5 | 20-50 | Branched pores | - | 5 | [65] |
| | Pulse and pulse reverse anodization | 18-25 | -7 to 30 | 30-60 | - | 250 | [55] |
| | Two-step anodization | 17 | 40 | 52 | 100 | 400 | [47] |

|  | Two-step anodization | 3 | 40 | 24 | 105 | - | [15] |
| --- | --- | --- | --- | --- | --- | --- | --- |
| **Sulphuric acid** | Two-step anodization | 1 | 15-25 | 18-26 | 49-65 | 250 | [56] |
|  | Pulse anodization | 1 | 25-35 | 16-35 | - | - | [57] |
|  | Hard anodization | 1 | 27-80 | 15-30 | 70-145 | - | [58] |
|  | Two-step anodization | -8 | 15-25 | 14-11.5 | 35-37 | 250 |  |
|  |  | 1 |  | 5.7-4.9 | 3.7-2.7 |  | [54] |
|  |  | 10 |  | 13-15.3 | 35.7-34.6 |  |  |
|  | Two-step anodization | 0.85 | 25 | 33 | 63 | 400 | [47] |
| **Phosphoric Acid** | Cyclic anodization | -1 | 20-300 | 300-1200 | - | 5 | [50] |
|  | Two-step anodization | 0 | 100-160 | 85-140 | - | - | [59] |
|  | Step Anodization | 25 | 160-20 | 160-30 | 580-87 | - | [60] |
|  | Two-step anodization | 12 | 195 | 254 | 490 | 400 | [47] |
|  | Two-step anodization | -5 and 0 | 100-160 | >100 | 250-490 | - | [61] |
|  | Two-step anodization | 0 | 160-195 | 140-190 | 405 | - | [62] |

Bar graph for the effect of different anodization parameters on pore size in two-step anodization process with varied anodization parameters and resulting pore size reported in the literature is given in **Figure 9** [46,47,54,56–58,60,63,64,66–68].

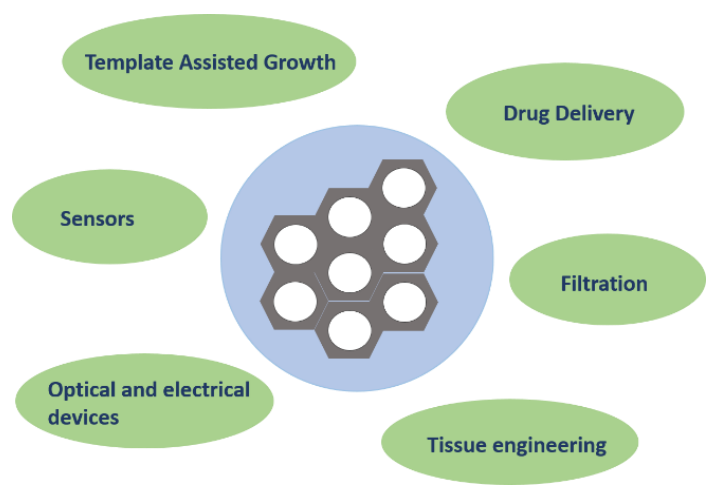

**Figure 9.** Applications of AAO.

## 5 Properties of AAO membranes

These membranes are highly flexible nanomaterials that consolidate the artificially steady and precisely vigorous properties of ceramics with a homogeneous nanoscale association that can be tuned in terms of pore distance and lengths. The AAO membranes have a high surface area that can accommodate atoms of diverse sizes. AAO membranes exhibit extraordinary properties with tunability which make them suitable for countless applications.

The AAO membranes are most commonly used for template-assisted growth of various nanostructures like nanotubes, nanorods, and nanowires of different organic and inorganic materials. Other applications include controlled release drug delivery, sensing applications, and size-based separation of materials in water filtration and hemodialysis. The properties responsible for great attention received by AAO membranes to get the maximum benefit from AAO membranes are discussed below.

### 5.1 Electrostatic properties

Electrostatic properties of the inward pore surface just as the neighborhood causticity inside the nanochannels can be tuned. The method used to manipulate the membrane features is electron paramagnetic reverberation (EPR) spectroscopy of a little particle – ionizable nitroxide – to get local acidity of nanochannels and local electrostatic potential in the quick region of the pore surface. Due to the amorphous nature of the alumina layer, the nanochannel surface of AAO membranes prepared by anodization shows a positive charge. EPR spectroscopy was used to study the electrostatic properties of the membrane. The EPR titration curve revealed that by annealing the membrane at a high temperature of about 1200°C, a phase transition occurs from amorphous AAO to crystalline AAO, and the surface charge of the AAO membrane is reversed. The anodization conditions greatly influence the surface electrostatic properties of the AAO membrane, and the acidity of the aqueous medium inside the pore channels depends on the pore diameter. Tunability of surface charge and associated electrostatic potentials make the membrane beneficial for specific applications that rely on acid-base equilibrium [69].

### 5.2 Wettability

The wettability of any droplet deposited on a solid surface depends on the energy difference of both liquid and the surface. In the case of the AAO membrane, energy difference governs whether the liquid will enter the nanopores or not. The diameter of the nanopore if increases, then ultimately, the contact angle of the water with the surface increases [70]. Electrochemical anodization of the aluminum substrate can result in superhydrophobic AAO membranes if modified with fluorinated silane solution. The hydrophobic membrane shows long-term stability and can be used in anti-corrosion membrane applications. The contact angle of ~165 and sliding angle < 2 have been reported for membranes modified by fluorinated saline [71]. Also, hydrophilic AAO nanoporous membranes can be prepared using the anodization method. By varying the anodization and the post-processing conditions, AAO membranes with different wettability and contact angles for water can be prepared depending on the application for which it is to be used [72,73].

### 5.3 Mechanical properties

The nanoindentation method is used to investigate the mechanical properties like surface hardness, scratch hardness, and surface roughness of the AAO membrane. The hardness of the membrane depends on the pore diameter; as the pore diameter increases, the hardness decreases. Heat treatment after membrane formation influences the mechanical properties of the membrane, hardness increases as a result of heat treatment. Also, the wettability of the membrane depends on the surface roughness, and smooth surfaces produce low surface energy [74,75]. Relatively thin AAO membranes with increased mechanical strength have also been reported by using Si support. Which results in better performance of membranes in biomedical applications [76].

### 5.4 Biocompatibility

AAO nanopores display excellent biocompatibility towards cells of the four basic tissue types (neuronal, epithelial, muscle, and connective tissue) as well as with platelets and different microorganisms. AAO interfacial examinations, a tremendous scope of biomedical applications have arisen. Nonporous AAO membranes have just been joined into coculture substrates for tissue designing, alumina biosensors, what's more, bone embed coatings or nanoporous bio capsules for drug conveyance. Some portion of these applications has likewise been concentrated in vivo in brief timeframe tests yielding promising results concerning biocompatibility, drug discharge properties, what's more, mechanical security of the nanoporous AAO films [77]. Biocompatibility of AAO films with a mean pore measurement of 100 nm was studied. A natural assessment of the layers was finalized to decide cell grip and morphology utilizing the Cercopithecus aethiops kidney epithelial cell line. Examination's positive results show that AAO films can be utilized as a feasible tissue platform for potential tissue designing applications later on [78].

## 6 AAO membrane applications

AAO is an electro-synthetic process that changes the surface of the metal through oxidation to deliver an anodic oxide layer. During this process, a self-coordinated, exceptionally demanded cluster of round and hollow formed pores can be made with controllable pore widths, periodicity, and thickness. These properties engage AAO layers to be used as formats in various nanotechnology applications without the necessity for expensive lithographical systems, and few applications are given in **Figure 9**. The current status of the investigation on AAO for various nanotechnology application make them beneficial in the assembly of nanomaterials and devices or unite them into unequivocal applications, for example, organic/substance sensors, nano-electronic gadgets, channel layers and clinical platforms for tissue designing [40].

Properties of AAO membranes with Various reported applications in different fields of interest are given.

## 6.1 AAO membrane for template-assisted growth of nanostructures

### 6.1.1 Applications of AAO membrane for template-assisted growth of nanostructures

Template-assisted growth represents a straightforward method in which the template acts as a structural framework. The nanostructures of the desired material can be formed within the template from precursors in situ, with shape and size defined by the pore size of the template. Generally, templates are of two types; soft templates and hard templates. Naturally occurring micelles are referred to as soft templates, whereas templates related to materials like AAO membranes are considered hard templates [79]. Due to size, the influence of diminishment nanostructures have a higher active part and increased surface to volume ratio, in addition, low-cost synthesis, monodispersed, and controllable diameter of AAO nanoporous membranes makes them suitable for the template-assisted growth of different nanostructures. Nanostructures ranging from nanotubes, nanowires, nanorods, nano dots arrays, interconnected 3D nano networks with highly controllable size and shape of a certain material can be grown using such membranes [80,81] Successive deposition of specified material into the nanosized pores of the AAO membrane results in nanostructures embedded in the membrane. However, free nanostructures can also be obtained by removing the membrane acting as a template [82,83].

AAO template-based fabrication of nanowires arrays of pure metals like Pb [84], Sb [85], Ni [86] [87], Zn [87] , Ag [87] has been reported using electrodeposition. S. Thongmee et al. [88] also used electrodeposition to fabricate metallic nanowires of Ni, Co, Cu, and Fe. Rui Yang et al. [89] reported low ion concentration and heating rate beneficial for the AAO template-based fabrication of Ag nanowires using calcination. For the first time, Zhao Wu [90] et al. used AC-DC electrodeposition for deposition of Au nanowires into an AAO template. Besides, the galvanic displacement method has been employed by Arulkumar Ganapathi et al. [91] to synthesize Cu nanowires into AAO. Electrodeposition method has also been employed for the fabrication of Ag dendrites to be used as 3D Surface Enhanced Raman Spectroscopy (SERS) active substrates for the detection of trace organic pollutants [92]. **Figure 10** depicts step-wise fabrication of Au nanoparticles on AAO template. Other nanostructures of different materials reported in literature formed using AAO template along with their applications are given in **Table 4** [88–91,93–109].

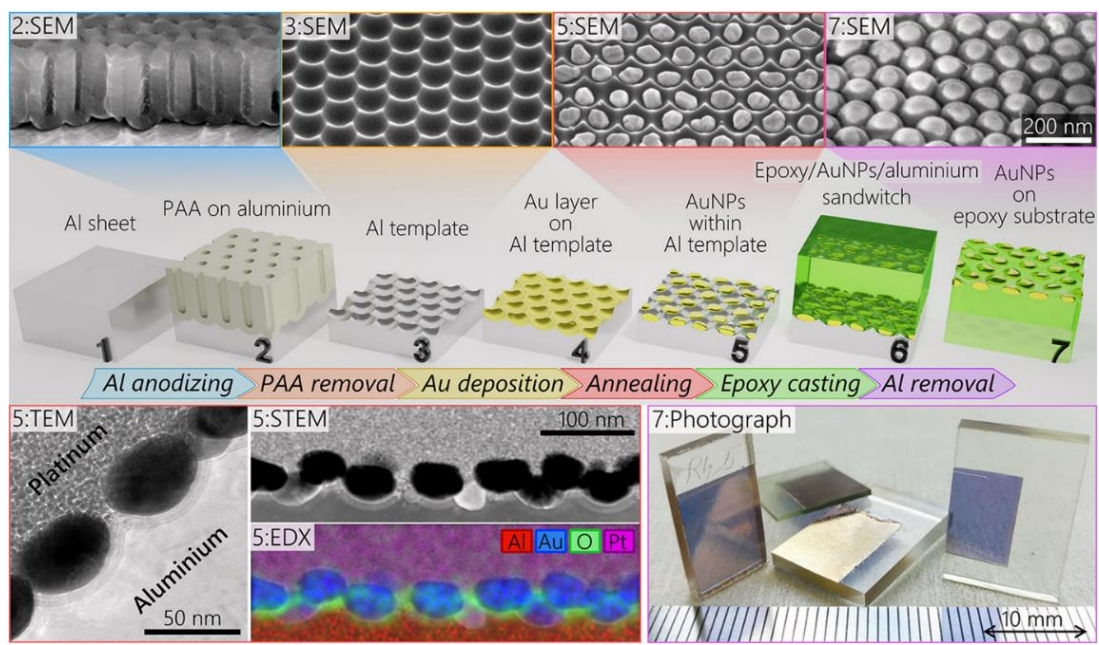

**Figure 10.** AAO based fabrication of AuNps on epoxy substrate for label-Free Plasmonic DNA Biosensors. Step wise schematic illustration of fabrication process (top), SEM figures of membrane at different steps (center), TEM STEM and EDS of AuNps within Al template (bottom left), photograph of sample after 7 steps (bottom right). Reproduced from [103]. Copyright 2020 American Chemical Society.

**Table 4.** AAO membranes for template assisted growth of various nanostructures.

| Pore size of AAO (nm) | Method | Nanostructure formed | Diameter of nanostructure (nm) | Application | Refs. |
|---|---|---|---|---|---|
| 100 | Calcination | Ag nanowires | 35.1 | Potential application as a catalyst | [89] |
| 50 | Electrodeposition | Cu nanowires | 50 | Single crystalline nanowire formation with good magnetic properties | [88] |
| 30 | AC-DC electrodeposition | Au nanowires | 30 | - | [90] |
| 67<br>110 | RF magnetron Sputtering | Au nanoparticles | 55<br>92 | Label free plasmonic DNA biosensos | [103] |
| 80 | Electropolymerization | Poly(3-hexylthiophene) (P3HT) nanorods and nanotubes | 100 | Effect of monomer concentration on nanostructure and corresponding band gap | [104] |
| 50 | Pulsed laser deposition | $CoFe_2O_4$ nanodots | 45 | Potential in oxide nanomagnets and spintronics | [105] |
| 60 | Microwave plasma electron cyclotron resonance chemical vapor deposition (ECR-CVD) | Carbon nanotubes | 75 | potential applications for cold-cathode flat panel displays | [106] |
| 75 | E-beam evaporation | ZnO nanodots on Si(100) | 9.7 | - | [107] |
| 78 | Galvanic Displacement process | Cu nanowires | 73 | Electrochemical denitrification | [91] |
| 80 | Infiltration and sulfurization | $CuInS_2$ nanorods | 80 | Potential for fabrication of Photocathodes | [108] |
| 200 | CVD | Carbon nanocoils | 100-500 | - | [109] |
| 80 | Hydrothermal process in aqueous solution | ZnO nanowires | 80 | Field Emission Applications | [93] |
| 300 | Plasma assisted reactive evaporation | Tree like structures of InN nanoparticles | Thickness 500 | - | [94] |
| 30 | Electrochemical deposition | ZnO nanorods | 60 | Fast Response Photodetector | [95] |
| 230-370<br>40-100 | Sol gel routes | $\gamma$-$Na_{0.7}CoO_2$ nanotubules | 200-340<br>35-97 | Thermoelectric applications | [96] |
| 60-70 | Electrodeposition | Co nanowires | 70 | Catalyst for methane combustion | [97] |
| 90 | Ultrasonic hydrothermal method | $Fe_3O_4$ nanoparticles | 30-40 | Arsenic removal | [98] |
| 100 | Electrodeposition | Cu nanowires | 100 | Antibacterial activity | [99] |
| 60 | Etching | Aluminium oxide nanowires | 67 | Microelectronic devices | [100] |
| 240 | Electrodeposition | ZnO nanowires | 300 | Scintillator | [101] |

| 75 | Autoclave enclosed hydrothermal growth | Zr-BiFeO$_3$ nanoflakes | 25-100 | Energy storage application | [102] |
| 51 | Solid state dewetting | Au nanoparticles | 51 | Plasmonic DNA bosensors | [103] |

*6.1.2 Prospects of AAO membrane for template-assisted growth of nanostructures*

AAO membranes have excellent prospects as a template for the growth of large-scale nanostructure arrays for practical applications. The template-assisted growth of nanostructure is a widely accepted theoretical concept, but studies are required for the subset of materials to develop various structures. Optimization of template synthesis is crucial for the growth of multidimensional structures on a large scale. The control over crystallinity, diameter, and functionalization of nanostructures fabricated using AAO templates may open new paths for the expansion of their applications in various fields.

**6.2  AAO membrane for filtration**

*6.2.1  Applications of AAO membrane for filtration*

Pressure driven membrane filtration process depends upon the pore diameter of the membrane and is divided into four types. The factor for good selectivity is that the membrane must have narrow pore distribution. The thickness of the membrane should be between 1 to 10μm for high flux. The size-based separation using membrane is used in many industrial and medical applications. The membranes with a mean pore diameter of 1μm are used in the filtration process called microfiltration. Ultrafiltration is performed using a membrane with a mean pore diameter of 100nm. For nanofiltration, the membrane should have a pore diameter of 10nm, and the membranes with 1nm mean pores are used in the process called reverse osmosis.

Biofiltration can be categorized into two main mechanisms based on the size and chemical affiliation of filtration membranes. Size limitation results in the blockage of molecules whose size is greater than the membrane's pores. Chemical affiliation is related to the filtration mechanism as in the bonding of molecules with the filter membrane, and the selected molecules can be separated based on lower or higher chemical affinity with the membranes. If these factors can be controlled, then the filtration process can be optimized easily [110–113]. The simplest membrane filtration system comprises of a filtration membrane set up in the system, fluid to be filtered is poured onto the membrane, and the filtrate is obtained.

The separation of desired molecules has applications in the biomedical field. Important factors for filtration application of AAO membrane include flux rate, selectivity, high chemical, and mechanical stability with uniform pore size. The porosity of the sheet and membrane thickness is used to determine the flow rate and produce a mechanically stable AAO membrane for biomedical applications. Theoretical studies have been done for testing the mechanical stability of AAO membranes for filtration application. The studies show that the membrane can bear pressure up to 0.79 MPa and is suitable for microdialysis [114]. An adaptive ultrafiltration layer has been designed using an AAO template in polyprotic to provide filtration properties and used for desoxyribonucleic acid analysis [115]. Two-layered AAO has also been reported to speed up the hemodialysis and hemofiltration process [116]. Two-layered AAO membrane structures have been used for effective dialysis and to sieve exosomes, potential biomarkers for the early detection of cancerous cells [117]. Functionalized AAO membranes for purification of contaminated water and enantioselective separation have also been reported in the literature [118,119]. **Figure 11** and **Figure 12** represent the mechanism of filtration using functionalized AAO membranes for filtration and separation respectively. Various other applications of AAO membranes for filtration purposes have been reported in the literature and are also given in **Table 5** [110,113,116,117,120–132].

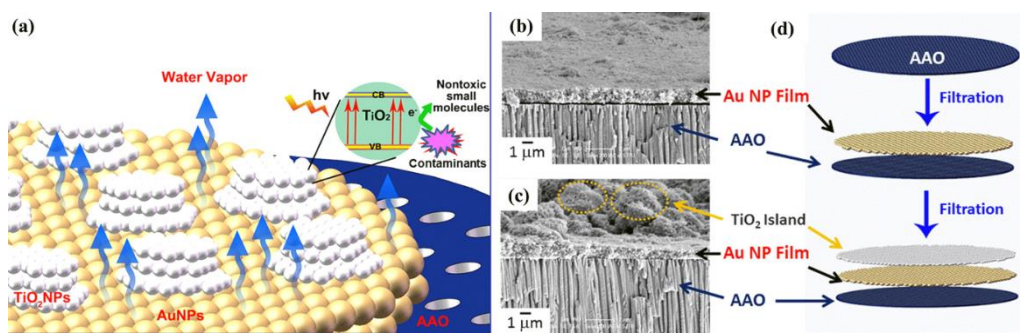

**Figure 11.** Biofunctional membrane for purification of contaminated water. (a) Design of Bifunctional TiO$_2$−Au-AAO membrane with AAO at bottom Au in the middle and TiO$_2$ at top, (b) Cross sectional SEM image of two layered structure (Au Np/AAO), (c) Cross sectional SEM image of three layered structure (TiO$_2$/AuNp/AAO), (d) Schematic illustration of preparation method. Reproduced from [118]. Copyright 2015 American chemical Society.

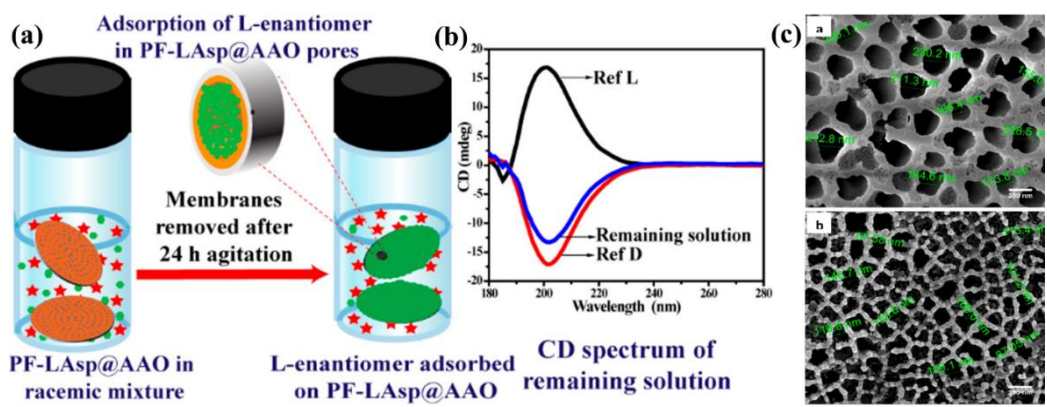

**Figure 12.** Functionalized AAO membrane for enantioselective separation. (a) Illustration of enantiomer separation process using Chiral Amino Acid Functionalized Polyfluorene Coated AAO (PF-Lasp@AAO), (b) CD spectrum of reamaining solution after separation process, (c) FE-SEM micrographs of AAO membrane and polymerized AAO membrane. Reproduced from [119]. Copyright 2020 American Chemical Society.

**Table 5.** Different types of AAO membranes for filtration.

| Membrane type | Pore size of AAO membrane (nm) | Application | Refs. |
| --- | --- | --- | --- |
| AAO based 3D microchannels | 50 | Microchannels for molecular separation | [127] |
| Two nano-porous AAO membranes sandwiched in between PDMS layers | 30<br>80 | Two AAO membranes sandwiched between PDMS for Exosomes Isolation | [117] |
| AAO membrane in micro hydraulic mechanical system | 50 | Vibro-active nano filter for filtration, separation, and transportation of particles, using standing surface acoustic waves (SSAW) | [128] |
| Optical gas sensing package covered with AAO membrane | 40 | Filtration interface to let air and other gases pass and prevent contaminations in forms of particles for an optical gas sensor | [129] |
| AAO tubular membrane | 20 | Potential application for hemodialysis | [130] |
| AAO membrane and polydimethylsiloxane (PDMS) microwells with through holes | 25 | Microwell-assisted filtration of algal solution | [131] |
| Nafion-coated AAO filter | 20 | Nafion-coated AAO filter for ion separation in presence of electric field | [132] |
| Microfluidic channels of AAO with immobilized probe DNA sequences | 200 | Filtration membrane to form probe DNA electrode for electrochemical sensing | [120] |
| Polyrhodanine modified AAO | 150 | Heavy metal ions removal with potential applications for removing the hazardous heavy metal ions from wastewater. | [121] |
| Al textured AAO (AT-AAO) membrane | 31.25 | Fluid flow and permeability of acetone, methanol, dimethylformamide, water, cyclohexane, ethanol, isopropyl alcohol, and n-butanol | [122] |
| Polyrhodanine modified AAO with activated carbon deposition | 55-70 | Multi-filtration membranes with potential applications in removing heavy metal ions from water, soil, food and drug | [123] |
| Two layered AAO membrane structure | 50 350 | Wearable hemodialysis device | [116] |
| AAO membrane integrated with an on-chip microfluidic platform | 38.2 | Electrokinetic separation of biomolecules and potential for binary separation and detection of a molecular reaction as a result of specific recognition | [124] |
| Functionalized nanoparticles embedded AAO | 100 | As working template for enhanced filtration | [125] |
| Silica–surfactant nanocomposite in the AAO membrane | 3.5 | Size-based molecule separation | [113] |
| Functionalized chitosan AAO membranes | 200 | Purification of Hemoglobin from red cell lysate | [110] |
| Tubular AAO films | 60 | Liquid or gas filters, drug delivery, and energy applications | [126] |

*6.2.2 Prospects of AAO membrane for filtration*

The AAO membranes have promising prospects in gases and liquids filtration applications. Chemical affinity is an essential factor, and functionalization of AAO membrane with various nanostructures are important factors to be studied for enhanced and optimized filtration. Research on the bonding of multiple molecules with the membrane and decoration of nanostructures within pores can lead to efficient and highly selective filtration. The improved filtration performance may be proved highly promising to be applied in future areas.

**6.3 AAO membrane in sensors**

*6.3.1 Applications of AAO membrane in sensors*

Due to its unique properties, extensive studies have been performed on AAO to develop low-cost sensing devices. The characteristic responses of AAO membranes while interacting with light make them suitable for optically active devices. Effectively immobilized sensing elements within the pores can optimally interact with the analytes flowing through the pores due to the large surface-to-volume ratio of the nanopores. Also, AAO with nanoporous structure has potential in single-molecule sensitivity. These properties make them suitable for the development of electrochemical and biosensors. Nanopores of AAO membranes are suitable for molecular transport and attachment, resulting in one device integrated with both separation and sensing functions. Additionally, AAO membranes have also been effectively used as a substrate to fabricate various humidity and gas sensors [133]. Highlights of sensing and biosensing applications of AAO membranes present in the literature are given.

G. Gorokh et al. [134] prepared two distinct aluminum oxide films using tartaric and malonic acid. The anodic films were first prepared, then tungsten oxide was sputter deposited on the films to form metal oxide test sensors. The results indicated that the sensitivity of the sensors using tartaric acid increases with the increased porosity of the membrane. Both the test sensors were unresponsive for carbon monoxide but showed an enhanced response for ammonia. Ning Hang et al. [135] deposited metal oxide gas sensing film of Ga-doped ZnO (GZO), and GZO annealed in $H_2$ (GZO-H) via electrophoretic deposition on AAO/Al structure. In the presence of aqueous solution, the substrate and deposited film were destroyed because of the $H_2$. On the contrary, the deposition was successful in the presence of ethanol as a dispersant. In gas sensing AAO layer act as an insulator, so its thickness plays a vital role in enhancing its resistance. Formaldehyde was used as a probe to verify the gas sensing property of the as-prepared structure. It was concluded that AAO/Al structure could be used to fabricate metal oxide gas sensors successfully. Pei-Hsuan Lo et al. [136] used anodic aluminum oxide with nanopores of about 50-60nm to implement inductive proximity sensors. The anodic aluminum oxide was used as a template for patterning spiral film of gold. Due to nanotexture, the gold film has significantly enhanced surface area that improved the inductance of the proximity sensor. Chitsung Hong et al. [137] developed a metal-insulator-metal device in which the nanoporous AAO fabricated on the Si substrate act as an insulator. Au-AAO-Al layers on the Si substrate were implemented. The top metal layer deposited on the AAO has nanotexture, due to which the capacitive touch sensor has improved sensitivity. AAO based humidity capacitive sensors were investigated by Mamadou Balde et al. [138] Paper was used as a substrate to fabricate the AAO layer. The results showed clearly that the geometry of AAO and the sensitivity of the sensors varies by voltage variation. The capacitive response of the sensor increased with temperature up to $45^0C$ but decreased for measurements after two weeks. Also, the low sensitivity of AAO based capacitive humidity sensors under low humidity conditions has been addressed by H. Park et al. [139] The AAO barrier oxide layer stored humidity within the pore and resulted in linear capacitance variation at relatively low humidity

Tushar Kumeria et al. [140] integrated AAO substrates coated with ultrathin gold layer into microfluidic devices. Two different strategies were used in this work to explore sensing properties based on reflective interferometric spectroscopy. Gold-modified AAO with Thiol binding and antibody-modified AAO with circulating tumor cells binding were studied. Results showed potential for biosensing devices with the admirable capability of detecting different analytes. Besides, Mohamed Shaban et al. [80] fabricated nanoporous CdS/AOA bi-layer film using AAO as a template for spin coating of CdS film. Due to the positive experimental results, the bi-layer film was attributed as an efficient, low-cost, and relatively simple glucose biosensor. Label-free DNA detection is the most challenging application of Localized surface plasmon resonance sensors due to the petite size of the DNA molecules. Tomas Lednicky et al. [103] fabricated gold nanoparticles-epoxy surface nanocomposites using AAO as a template. The fabricated LSPR sensors were accurate in the detection of 20 bases long label-free DNA. Ma et al. [129] fabricated optical gas sensor with an AAO membrane as the air filter. The fabricated system could prevent contamination with particles above the size of 70nm. However, the sensor's sensitivity could be improved by tuning the pore size of the AAO membrane. Schematic representation for fabrication of a pressure sensor using AAO membrane with its current response under pressure is provided in **Figure 13**.

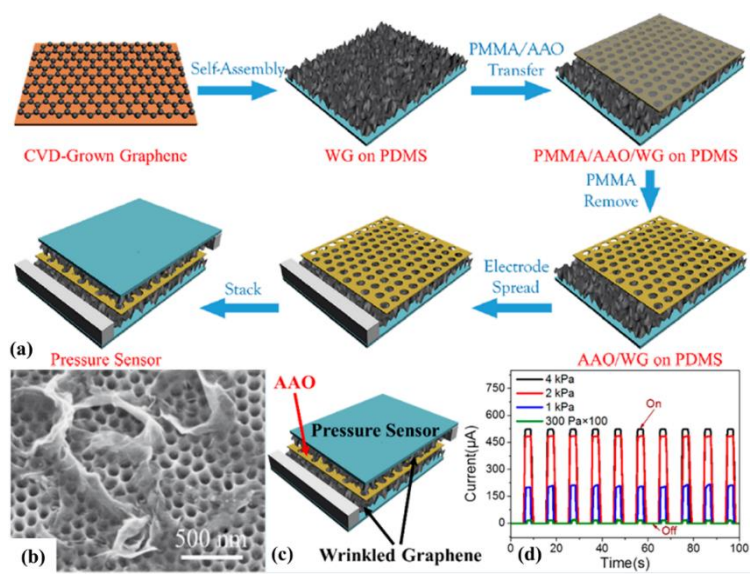

**Figure 13.** Ultrathin Pressure Sensors AAO membrane for insulation of two graphene layers. (a) Schematic representation of steps for fabrication process of pressure sensor, (b) SEM image of WG on AAO membrane, (c) Current response under various pressure. Reproduced from [141]. Copyright 2017 American Chemical Society.

Few notable applications of AAO in sensing are given in **Table 6** [103,117,120,129,134,137–139,141–151].

**Table 6.** Different types of AAO membranes for sensing applications.

| Membrane | Pore size (nm) | Effect of pore size | Sensor type | Application | Refs. |
|---|---|---|---|---|---|
| Au/AAO/Au | 35, 100 | No significant effect of the pore diameter was observed | Capacitive sensor | Sensitive to He, $SF_6$, $CO_2$, $N_2$ and also applicable for sensing wide range of gases/vapors | [151] |
| AAO membranes as dust and moisture filtration interface | 40, 80 | For increased pore size higher diffusion factor but a lower filtering performance | Optical gas sensor | $CO_2$ gas sensing and potential for other types of gas sensing | [129] |
| AAO membrane as the binder | 30, 300 | Response time reduces as the pore diameter increases | Pressure-sensitive paint | Suitable for developing multi-layer paint systems | [142] |
| WG/AAO/WG | 80-90 | - | Pressure sensor | High-performance pressure sensor is promising for integration in flexible electronic devices to fulfill tactile sensing | [141] |
| Ag NPs immobilized AAO membrane. | 20 | - | Surface enhanced Raman scattering (SERS) sensing | Potential for a multitude of applications such as process monitoring in energy generation and production, molecular adsorption and desorption during high temperature catalysis involving precious metals | [143] |
| Two nano-porous AAO membranes was designed to sieve exosomes | 80, 30 | Size of exosomes is 30–100 nm in diameter porse size greater than 100nm is not effective. | Isolation and quantitative analysis of exosomes. | Early cancer detection | [117] |
| AAO filtration membranes with microfluidic channels | 200 | - | 3-D microfluidic-channel-based electrochemical DNA biosensor | Flow-forced filtration hybridization | [120] |
| AAO layers on paper | 40-90, 40-150 | Significant increase in capacitance at high humidity for sensor with larger pore size | Humidity sensors | With some treatment stable capacitance response can be obtained. | [138] |
| Multiple AAOs by simultaneous multi-surfaces anodization (SMSA) | 24.5, 31.3, 37.3, 44.2, 44.9, 45.2 | - | Humidity sensor | Templates and sensing applications | [139] |
| AAO as template for | 67, 110 | Pore size determines NPs | Plasmonic sensor | Label-free detection of DNA | [103] |
| | | | | arrangement and interparticle distances | |
| AAO nanopore patterns inside the Fabry-Perot interferometer (FPI) cavity | 35 | - | Optical biosensor | Transparent nanostructured FPI device for highly multiplexed, label-free biodetection | [144] |
| Two metal films with AAO as dielectric layer | 70 | - | Capacitive-type touch sensor | Detection of small object such as Drosophila | [137] |
| Silver/AAO arrays | 70 | - | SERS sensors | SERS detections and many applications based on plasmon induced by 2-D silver nanoparticles encapsulated by AAO | [145] |
| Gold nanowire (GNW)/AAO | 90 | - | Metal ion sensor | Sensing electrode for applications in biochemistry and electrochemistry | [146] |
| AAO template | 400 | - | Triboelectric nanogenerator (TENG) Self-powered sensitive sensor | Real-time mobile healthcare services for the management of chronic diseases | [147] |
| AAO-assisted $MoS_2$ honeycomb structure | 200 | - | Humidity Sensor | Wearable sensor for multifunctional applications such as noncontact sensation of human fingertips, human breath, speech recognition, and regional sweat rate | [148] |
| AAO photonic crystals | 30-40 | - | Concentration sensor | Enhanced photoluminescence for optical devices | [149] |
| $WO_3$ deposited AAO | 60 | - | Metal oxide sensor | potential utilizations in the fields of energy storage devices, micro-, opto and nanoelectronics. | [134] |
| $SiO_2$ coated AAO | 74<br>65<br>61 | Greater aqueous stability for smaller pore size | Label-free optical biosensing | Applicable for highly sensitive and specific biomolecular interaction detection | [150] |

### 6.3.2 *Prospects of AAO membrane in sensors*

AAO membranes have promising prospects in sensing applications. High sensitivity and outstanding sensing performance of AAO membranes have been reported. Further progress in superior sensing performance is expected by structural improvement and chemical modification. AAO based implantable biosensors can be explored to be used as immunosensors for biological systems monitoring. So, miniaturized and integrated lab on chip systems based on AAO membranes may further develop sensing applications. For further advances in AAO based sensing and biosensing devices, bio-diagnostics and environmental toxic agent detection are the main focus areas.

### 6.4 AAO membrane in drug delivery

### 6.4.1 *Applications of AAO membrane in drug delivery*

A drug delivery system aims to deliver a definite amount of medicine to a specific location at a precise time [152,153]. This method helps in the targeted medication delivery, but the system's main challenges are bypassing hepatic metabolism and constant drug release to avoid multiple doses [154]. Drug delivery systems can be categorized as osmotic and diffusion-controlled membrane systems. The osmotic system contains a semi-permeable polymeric membrane that is permeable to water but not permeable to the drug. While diffusion-controlled membranes systems work on the diffusion principle, the transport of the drug across the membrane and the thickness of the membrane determines the drug release. In this type of system, porous as well as non-porous membranes can be used [9].

The properties like chemical stability and the possibility of functionalization of membrane surface for enhanced targeted drug delivery make anodic aluminum oxide membrane an excellent candidate to be used as a drug delivery system [81]. The intracellular behavior of tubular AAO structure has been investigated to develop new pathways for AAO nanotubes to be used as drug carriers for the targeted delivery of drugs[155]. In nanomedicine, nanomaterials have gained much attention from researchers to establish reliable drug delivery systems [156]. AAO membrane can also be effectively used to form hollow nanostructures, these hollow nanostructures can be used as drug carriers after the encapsulation of drugs [157–160]. Piezoelectric nanorobots referred to as nanoeels using AAO templates have also been reported for targeted drug delivery. With the help of a magnetic field, these nanoeels could be steered to the target location for on-demand drug release [161].

Polymeric micelles as nanocarriers loaded in AAO membrane can also be used to deliver poorly soluble drugs. The deposition of ultrathin polymer film via the plasma polymerization method helps in the constant release of medication for an extended period [162]. Kristem E La Flamme et al. [163] used AAO biocapsules fabricated by anodization for controlled release of insulin for immunoisolation. the study proved that cell density and capsules transport area plays a vital role in the optimization of the output of the device. Ho-Jae Kang et al. [164] fabricated AAO on stents for controlled release of 2-deoxyadenosine. The effect of pore size and diameter on the drug release was studied. The results indicated that the amount of drug released during a specified time interval was directly proportional to the pore diameter and inversely proportional to the pore depth. Spomenka Simovic et al. [165] combined structural and chemical modification of the pores of AAO membrane by plasma polymer deposition. They proposed that the plasma polymer layer would help control the diameter of pores at the surface of the membrane hence resulting in controlled drug release. It was concluded that the plasma polymerization reduced pore size at the surface and helped modify the physical and chemical properties. This method provides enormous flexibility for fine-tuning the drug release. In a study [166], the effect pore diameter and pore depth of AAO membrane for drug release of Paclitex, a drug used in cancer chemotherapy. The results showed that the diameter of the pores does not affect. However, pore depth has a direct relation with drug release. Sung Bum Park et al. [167] used an AAO membrane to fabricate mesoporous silica nanorods for controlled release of the anti-cancer drug doxorubicin. The nanorods proved to be ideal for cancer treatment, and the addition of Na content to the nanorods results in increased biodegradation speed of the nanorods.

**Figure 14** presents a schematic of drug storage and release using aligned caron nanotubes embedded in AAO for the controlled release of drugs. Chitosan functionalized AAO membrane for the formation of drug release system with the schematic and experimental setup is provided in **Figure 15**.

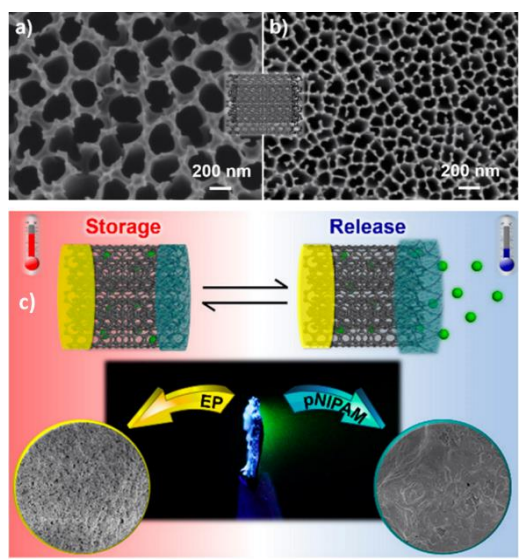

**Figure 14.** Modified VACNTs embedded in AAO for controlled drug release, SEM images of (a)top and (b) bottom parts of VACNTs inside the AAO membrane, (c) Schematic illustration of fluorescein storage and release mechanism from Electrophoretic paint (EP) at right and poly(N-isopropylacrylamide) (pNIPAM) at left and optical image of fluorescein release as a function of time at center. Reproduced from [168]. Copyright 2020 American Chemical Society.

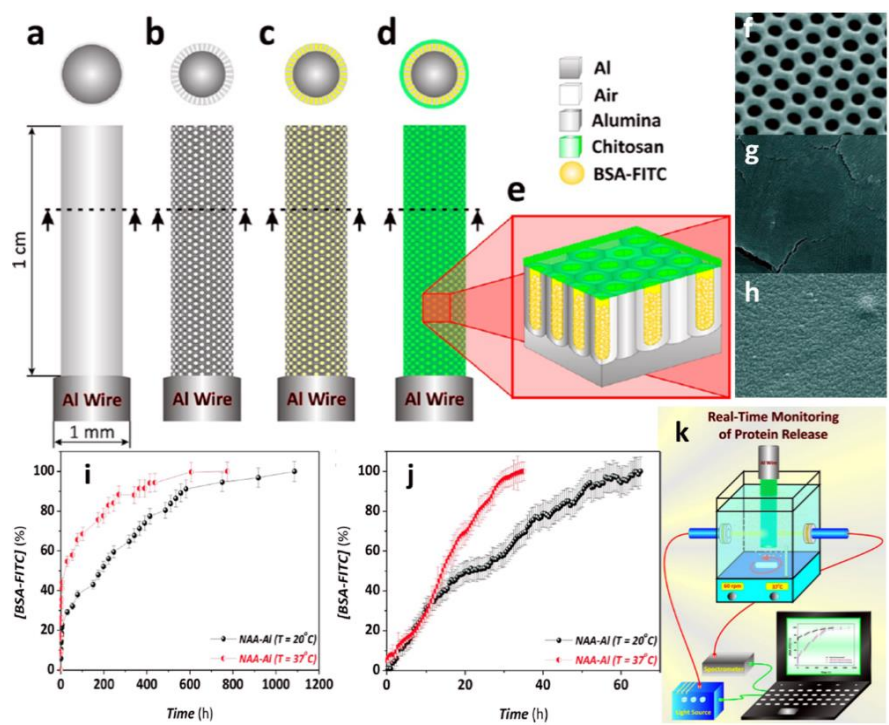

**Figure 15.** Therapeutics releasing AAO nanowires. Schematic representation of proposed therapeutic releasing system formation (a) electopolished Al wire, (b) after anodization (c) After protein loading (d) after chitosan coating, (e) cross sectional view, (f) SEM of nanoporous AAO-Al wire after pore widening (scale bar = 250 nm), (g)Top view of NAA-Al wire after 1 cycle of chitosan coating) (scale bar = 1 μm), (h) Top view of Nanoporous AAO-Al wire after 5 cycles of chitosan coating (scale bar = 1μm), Effect of temperature on protein release performance from chitosan-coated NAA-Al wires under static (i) and dynamic conditions (j), (k)Schematic of experimental setup to check therapeutic releasing performance. Reproduced from [169]. Copyright 2015 American Chemical Society.

Various applications of AAO membrane in drug delivery systems are given in **Table 7** [111,162,165–179].

**Table 7.** AAO membranes for drug delivery

| Membrane type | Pore diameter (nm) | Pore length (μm) | Drug loaded | Efficiency/ behavior of drug release | Application | Refs. |
|---|---|---|---|---|---|---|
| Plasma polymerized nanotubular AAO membrane | 65-160 | 20 | Indomethacin and fluorescent dye loaded in polymeric micelles | 31-55% drug release in 6-8 hours followed by the slow release over 8 to 22 days. | Porous therapeutic implants for an extended elution time. | [162] |
| Electrically responsive AAO membrane electropolymerized by polypyrrole doped with the dodecylbenzenesulfonate anions (PPy/DBS) | 410 | 60 | Fluorescein isothiocyanate labelled Bovine serum albumin molecules (FITC-BSA) in 0.1 M NaDBS aqueous solution | On demand drug release with switching time less than 10s | Pulsatile drug release for metabolic syndrome and hormone related disease. | [177] |
| AAO with embedded silica nanotubes | 20 | 30 | A racemic mixture of the RS and SR enantiomers was used as feed solution. | 30nanomoles/15h | Enantiomeric drug separation. | [111] |

| Structure | Pore diameter (nm) | Thickness (µm) | Drug | Release | Application | Ref |
|---|---|---|---|---|---|---|
| PGMA nanotubes embedded inside AAO nanopores | 200 | 60 | Doxorubicin (DOX) | 67.47% and 43.66% released after 11 h for long and fragmented NTs respectively. | Diagnostics and Therapeutics for cancer | [178] |
| Silica deposited AAO template | 123 | 4 | Prepolymer solution with DOX. | - | Composite silica nanotest tubes for targeted cancer therapy | [179] |
| AAO nanotube structure | 20 | 38 | 1 ml of 1% amoxicillin in phosphate buffered solution (PBS) was loaded. | Burst effect in the first hour with ~13 µg of drug release with sustained release of 2 µg for 5 weeks. | Potential for therapeutic surface coatings on medical implants | [170] |
| Silicone elastomer polydimethylsiloxane (PDMS)-based contact lens with AAO thin films in central region | 140 | 2 | Timolol was mixed with a fluorescein dye loaded into the nanopores of AAO. | 90% of the drug was released after 30 days. | Contact lens device for Glaucoma diagnostics and in situ drug delivery | [171] |
| Nanoporous AAO membrane | 70 | 10 | Se with chitosan and indomethacin. | Cumulative mass release of 0.16mg after 300 hours | Promising alternatives for localized delivery of Se using simple and low-cost drug-releasing implants. | [172] |
| AAO as template for mesoporous silica nanorods | 200 | 50 | DOX loaded in silica nanorods | Initial burst was observed within the first 24 h and was linearly released (40µg) until the 8th day of incubation. | Anti-Cancer | [167] |
| TiO$_2$ coated AAO | 40-50 | 0.5 | Paclitaxel was loaded in AAO | 23µg released in 8 days with linear increase | Cancer chemotherapy | [166] |
| Plasma polymerized AAO | 80-90 | 20 | Vancomycin was used as a model drug | ~100% in 500 hours | Fine-tuned drug release platforms to various types of drugs and applications | [165] |
| Aluminum wires featuring nanoporous anodic alumina layers and chitosan coatings. | 50 | 62 | BSA-FITC is selected as a model drug. | Sustained release performance for up to 6.5 weeks | Potential platform for future clinical therapies based on localized release of therapeutics | [169] |
| Electrically gated AAO membrane | 80 | - | Ethacrynic acid and timolol maleate | 250µg/cm$^2$h ethacrynic acid at 2V 40µg/cm$^2$h timolol maleate at 2V | Potential for treatment of ocular diseases including glaucoma | [173] |

| Functionalized AAO with hydrophilic and hydrophobic agents | 80 | - | Indomethacin | Initial "burst" effect up to 65-75 % of cumulative release in the first 3 h with sustained release upto 120 h | Excellent platform for drug delivery applications | [174] |
|---|---|---|---|---|---|---|
| Fragmented AAO membrane | 130 | - | AAO particles embedded in a polymer matrix as the fillers loaded with Ag NPs | Elution of 50 hours without and 350 hours with PMMA sealing | Platform for in situ drug release | [175] |
| Functionalized AAO membrane with vertically aligned carbon nanotube (VACNTs) arrays | 200 | 60 | A fluorescent dye was used as a model compound | Fluorescent molecules can be stored in the VACNTs when the temperature is above 32 °C, and released when the temperature decreases below 32 °C | Drug storage and delivery | [168] |
| Electrospun coated AAO | 190 | 50 | Indole-3-acetic acid (IAAA) | 90% of IAA was released during the initial 5 h and a slow release till 24 h | Controlled drug release. | [176] |

*6.4.2    Prospects of AAO membrane in drug delivery*

AAO membranes have been used for effective drug delivery, but there is still plenty of room for advancement. Drug loading inside cavities of AAO membranes can be enhanced by emphasizing stable chemistries and modified anodization procedures. Furthermore, for successful drug delivery platforms, research related to the biocompatibility of the membrane to be used for delivering therapeutics and testing several medications could be crucial. The critical point is to redirect the published work for in vivo applications. For this purpose, the drug delivery system can be integrated with sensors or microchips for sustained and targeted drug delivery.

**6.5    AAO membrane in miscellaneous fields**

*6.5.1    Applications of AAO membrane in miscellaneous fields*

Highly variable dimensions make AAO capable of other significant applications includes optical, electrical devices and several applications in biomedical. Fabrication of liquid crystal display (LCD) panels using AAO membrane has been reported. The nanocavities of the AAO membrane were filled with liquid crystal molecules. Accurately aligned liquid crystals in AAO exhibited excellent performance to be used in an electro-optical device [180]. Gold nanoparticles deposited AAO membrane has been used to prepare plasmonic absorber with the ability to absorb a wide wavelength range. This plasmonic absorber can also be used in applications like solar steam generation [181].

E. Davoodi et al. [182] presented the idea of nonporous AAO usage in tissue culture, biofunctionalization, drug delivery, and biosensing. Cyclic and pulse anodization was used to get uniform pores. Moreover, AAO can also be combined with biomaterials to form a composite scaffold for greater advantage. Three AAO films having different pore sizes for application as nano topological apparatuses in disease research. Cell morphology and attributes were subject to the pore size of AAO layers. AAO-1 layers (∼22 nm) didn't prompt adjusted cell morphology, though AAO-2 and AAO-3 films were instigated around. Subsequently, permeable AAO films were helpful nano topological instruments for in vivo disease research because of their capacity to incorporate natural cell adhesion and growth on the anodized aluminum oxide membrane data, dependent on the cell microenvironments. The permeable AAO films like in vivo malignant growth conditions very well may be utilized for malignant growth target treatment [183].

Electric vehicles appear to be a reasonable contender to tackle the energy deficiency issue. The issues of the current lithium-particle battery are the low energy thickness restricted by the graphite anode. Also, the development of lithium dendrites that use lithium metal as an anode will prompt serious wellbeing issues. The utilization of AAO films has been proposed to improve the presentation and the security of lithium metal batteries [184]. Aluminum and its compounds are by and large used essentially for marine and clinical applications. An essential procedure to make polyaniline/chitosan/zinc stearate superhydrophobic coatings on aluminum with a small-scale nano surface structure by polymerization of aniline and sworn statement of chitosan and zinc stearate covering is also accounted for. The superhydrophobic surface shows the most important water-repellent property, which is liable for the antiadhesion of microorganisms. Also, incredible erosion obstruction of aluminum [185]. **Table 8** provides applications of AAO in miscellaneous fields [69,73,77,78,99,182–199].

**Table 8.** Miscellaneous applications of AAO membranes.

| Membrane | Properties Studied | Application | Refs. |
| --- | --- | --- | --- |
| AAO | Pore distance and length | Applications that rely upon corrosive base equilibria | [69] |
| AAO | Thickness | The upgraded erosion opposition | [199] |
| AAO | Cell adhesion and proliferation on AAO nanopore geometries and surface modifications | Nanostructured substrate for Cell development | [77] |
| AAO | Manipulation of chemical and structural features. | Tissue Replacement | [182] |
| AAO | Porosity and membrane thickness | Cell culture | [186] |
| AAO | Porosity | Ovarian malignant growth target treatment | [183] |
| AAO | Controllable pore breadths, periodicity and thickness | Tissue designing applications | [78] |
| AAO | Porosity and surface area | Biotechnology | [187] |
| AAO | Hardness, impact strength, tensile and wears resistance. | Automobile and aerospace | [188] |
| Cu nanowires on AAO | Resistive or capacitive properties | Electronic (manufacture of nanodevices) and medical services applications | [99] |
| AAO | Pore size | Inhibition of Lithium Dendrites by employing AAO membrane | [184] |
| AAO | Wettability of PAN/GO nanofibers | For Li-S batteries to improve the cycle steadiness and rate limit | [189] |
| AAO | Investigation of polymer coatings on the surface morphology and corrosion resistance. | Marine and clinical applications | [185] |
| AAO | Thickness | Power generation | [190] |
| AAO | Pore distances | High energy material science. | [191] |
| AAO | Reactor design, synthesis conditions, the catalytic role | Wall conductivity | [192] |
| AAO | Surface morphologies | For surface wetting | [73] |
| Nafion/AAO membrane | Membrane thickness | Composite proton exchange through-plane conductivity | [193] |
| AAO | Structure | High chemical and thermal conductivity. | [194] |
| AAO | Fiber structure | Water Sportage and through-plane conductivity | [195] |

| Cu pillars on AAO | Thermal conductivity | Electronic packaging technologies. | [196] |
| AAO sandwiched between zeolite imidazolate framework 8 (ZIF-8) and WO$_3$ | Rectification performance | Osmotic energy harvesting device | [197] |
| AAO with Carbon black nanoparticles | Light absorbance | Optoelectronic and integrated photonics | [198] |

Schematic illustrations of AAO membrane for applications in miscellaneous fields like electronic packaging technologies, energy harvesting, and optoelectronics are shown in **Figure 16**, **Figure 17,** and **Figure 18** respectively.

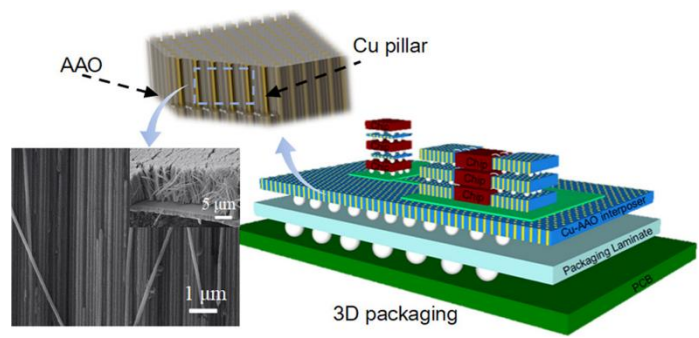

**Figure 16.** Schematic and SEM image of copper nanopillar array-filled AAO for anisotropic thermal conductive interconnectors. Reproduced from [196]. Copyright 2019 American Chemical Society.

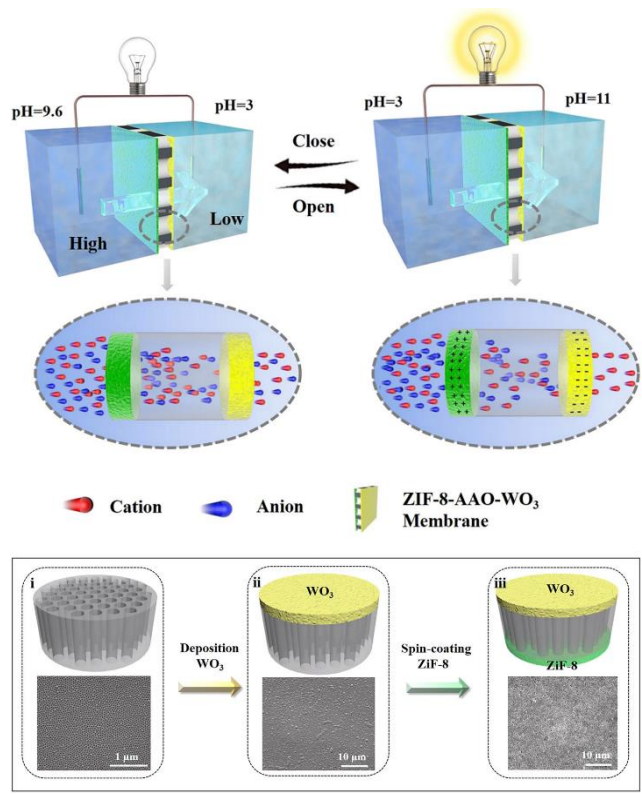

**Figure 17.** Schematic representation of the pH-driven Ion pool-structured multichannel gating membrane for the Osmotic energy harvesting device (top) , (i−iii) SEM images with step wise fabrication description of ion pool-structured nanofluidic diode with membranes of AAO, AAO−WO3, WO3−AAO−ZIF-8. Reproduced from [197]. Copyright 2021 American Chemical Society.

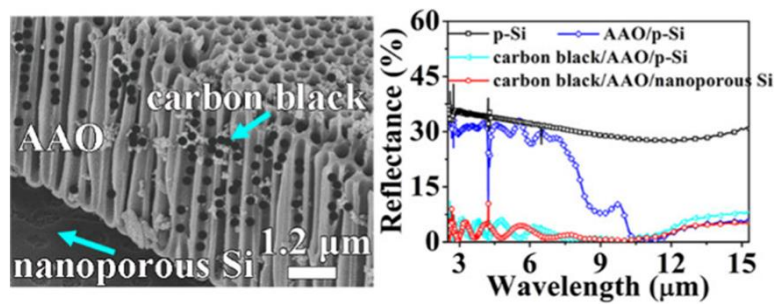

**Figure 18.** Carbon Black/AAO Templates on Nanoporous Si for optoelectronics and integrated photonics. SEM image of self assembled carbon black/AAO on Si (left), reflectivity of carbon black onto AAO templates on nanoporous Si (right). Reproduced from [198]. Copyright 2019 American Chemical Society.

### 6.5.2  *Prospects of AAO membrane in miscellaneous fields*

The prospects of AAO membrane applications translate into infinite possibilities of nanofabricated devices with enhanced properties and improved performance. Assembly of devices with enhanced properties devoted to specified application purposes, like energy and storage devices enhancing the storage capacity, can be achieved efficiently. The controllable features can be proved fruitful in the pharmaceutical, food, and energy industry. AAO based broadband absorbers can be used with other technologies for efficient thermal imaging and infrared detection. Nanochannels arrays pave the path for optimized nanofluidic diodes with applications in energy conversion zones.

## 7   Challenges of AAO membranes applications

Various challenges are associated with the application of AAO membranes. Two main types of technical challenges are worth highlighting: those related to evaluating the performance of devices and those related to scale-up under real-use conditions. Scaling up includes quality control of mass-produced nanoscale components, consistency of component integration, impurity management. Real-use suggests testing of the effect of deviations from prescribed operating procedures. There is a gap in knowledge related to the coating of nanomaterials with anodized aluminum oxide. Exposure and toxicity make it challenging to incorporate safety parameters into early decisions or designs. For the applications of AAO incorporated nanomaterials, the production of nanomaterials is expensive, and the characterization procedure of nanomaterials is also costly. Premium materials are required to ensure the reliability of applications of the nanomaterials embedded in the AAO membrane. Solar thermal applications need hard-anodized aluminum coatings, but the challenge in implementation is that AAO coating has higher rates; the higher budget is the major problem for producing any novel innovation for the extensive application area.

To examine AAO nanopores as cell interfaces for long-term in vivo response to multiple tissues is a challenge in developing future innovative biomedical devices that incorporate. AAO based nanosensors require comprehensive testing to confirm that they are reproducible and can perform accurately in the appropriate environment. An economic and business challenge in implementing AAO nanosensors is dealing with protection and intellectual property. Manipulation of the AAO membrane-based materials is complex on the engineering and nanoscale band-gap alignments, which is advantageous for separation, generation of charge, and transportation is worthy of intensive attention in the future. Another challenge is the selection of different functional materials and assembling them toward an integrated device. In energy storage and conversion applications. For energy storage and conversion applications, methods like an electrochemical deposition, physical vapor deposition, spin coating, and sol-gel infiltration are widespread for filling or covering the AAO template with active materials. The resulting materials are not usually single crystalline. So, it's a challenge how to improve crystal quality becomes critical to reducing charge-carrier loss in crystal imperfections. Comprehensive and systematic investigations are still required to thoroughly understand the transfer dynamics of photogenerated charge carriers in the complicated surface surroundings and explore a more effective means for harnessing solar energy. However, because basic science is constantly evolving, and patent granting organizations may lack adequate functional nanotechnology expertise to evaluate applications honestly, patents create considerable barriers to product commercialization in the field of nanotechnology.

The surface of the AAO membrane has rich content of hydroxyl groups that allow them to be easily modified with the help of organic molecules according to desirable functions. Despite all these techniques used for the modification of AAO, the application of functionalized AAO membranes will depend on the unpredictable needs of the market. Yet, these techniques only implement by the developments in the performance of the devices and cost. This area of science and development still holds different opportunities and challenges for researchers to investigate.

# 8 Conclusion and future directions

Porous membranes with desired features resulted in the development of membrane technology. Different methods have been used for the formation of such membranes. In this paper, we have presented an overview of different methods for porous membrane formation. Different anodization methods have been used to form AAO membranes and can be used to control the geometric features of the membranes by varying the parameters. The effect of different anodization parameters on the morphology of AAO porous membrane and its properties are presented. Empirical relations defining the geometric structure parameters of porous AAO have also been recognized. Cost-effective and self-organized anodization makes it attractive for high-tech industries and biomedical applications. Different applications reported in the literature in various fields are discussed, along with detailed applications. Following essential points can be deduced from applications of AAO.

- The applications of AAO strongly depend on the size and chemical affinity as the membrane parameters can be controlled easily.

- The AAO membranes can utilize all of the selectivity paradigms by modification using various molecules.

- Effectively immobilized sensing elements within the pores can optimally interact with the analytes flowing through the pores due to the large surface-to-volume ratio of the nanopores.

- Nanochannels of AAO promote the transfer of matter. In the presence of a barrier layer, the nanopores mimic nanocavities for containing medicine for targeted release or formation of various nanostructures with multiple applications.

- Proper functionalizing of the surface properties and defined geometry of porous AAO can further expand its applications, resulting in up-and-coming prospects for nanotechnology applications of porous AAO membranes.

Highly variable dimensions of AAO membranes make them capable of being used in many applications. However, depending on the unpredictable requirements, there are many challenges to overcome for AAO to be used commercially. More efforts are required to improve existing methods of functionalization of membrane and development of effectual techniques for the development of desired nanostructures using AAO. Enhanced stability and durability with improved methods may open new paths for promising applications.

## Author Contributions

**S. Manzoor:** Methodology, Writing- Original draft preparation; **M.W. Ashraf:** Data curation, Supervision; **S. Tayyaba:** Visualization, Validation; **M.K. Hossain:** Supervision, Writing- Reviewing and editing.

## Declaration of competing interest

The authors declare that they have no competing interests.

## References

[1] Michaels A. Membranes, membrane processes, and their applications: Needs, unsolved problems, and challenges of the 1990's. Desalination 1990;77:5–34. doi:10.1016/0011-9164(90)85018-6.

[2] CADOTTE JE. Evolution of Composite Reverse Osmosis Membranes. ACS Symp Ser 1985;269:273–294.

[3] J.Petersen R. Composite reverse osmosis and nanofiltration membranes. J Memb Sci 1993;83:81–150. doi:10.1016/0376-7388(93)80014-O.

[4] Norman N Li, Anthony G. Fane, W. S. Winston Ho TM. Advanced membrane technology and applications. John Wiley & Sons; 2011.

[5] C.ArnotDavideMattia KP. A review of reverse osmosis membrane materials for desalination—Development to date and future potential. J Memb Sci 2011;370:1–22. doi:10.1016/j.memsci.2010.12.036.

[6] Adiga SP, Jin C, Curtiss LA, Monteiro-Riviere NA, Narayan RJ. Nanoporous membranes for medical and biological applications. Wiley Interdiscip Rev Nanomedicine Nanobiotechnology 2009;1:568–81. doi:10.1002/wnan.50.

[7] Hossen S, Hossain MK, Basher MK, Mia MNH, Rahman MT, Uddin MJ. Smart nanocarrier-based drug delivery systems for cancer therapy and toxicity studies: A review. J Adv Res 2019;15:1–18. doi:10.1016/j.jare.2018.06.005.

[8]  Mahfuz AMU., Hossain MK, Khan MI, Hossain I, Anik MI. Chapter 2: Smart drug-delivery nanostructured systems for cancer therapy. In: Gil Gonçalves, editor. New trends smart nanostructured Biomater. Heal. Sci., Amsterdam, Netherlands: Elsevier; 2023.

[9]  Stamatialis DF, Papenburg BJ, Gironés M, Saiful S, Bettahalli SNM, Schmitmeier S, et al. Medical applications of membranes: Drug delivery, artificial organs and tissue engineering. J Memb Sci 2008;308:1–34. doi:10.1016/j.memsci.2007.09.059.

[10]  Bhadra M, Mitra S. Nanostructured membranes in analytical chemistry. TrAC Trends Anal Chem 2013;45:248–63. doi:10.1016/j.trac.2012.12.010.

[11]  V D. Review on Membrane Technology Applications in Food and Dairy Processing. J Appl Biotechnol Bioeng 2017;3. doi:10.15406/jabb.2017.03.00077.

[12]  ITAYA K, SUGAWARA S, ARAI K, SAITO S. Properties of porous anodic aluminum oxide films as membranes. J Chem Eng JAPAN 1984;17:514–20. doi:10.1252/jcej.17.514.

[13]  Leenaars AFM, Keizer K, Burggraaf AJ. The preparation and characterization of alumina membranes with ultra-fine pores. J Mater Sci 1984;19:1077–88. doi:10.1007/BF01120016.

[14]  Rehman AU, Ashraf MW, Tayyaba S, Bashir M, Wasim MF, Imran M. Synthesis and growth of bismuth ferrite ($BiFeO_3$) with lanthanum (La) and yttrium(Y) doped nano-structures on anodic aluminum oxide (AAO) template. Dig J Nanomater Biostructures 2021;16:231–8.

[15]  Ashraf MW, Manzoor S, Shahzad Sarfraz M, Wasim MF, Ali B, Akhlaq M, et al. Fabrication and fuzzy analysis of AAO membrane with manipulated pore diameter for applications in biotechnology. J Intell Fuzzy Syst 2020;38:5857–64. doi:10.3233/JIFS-179673.

[16]  Attiq-ur-rehman, Ashraf MW, Tayyaba S, Ali SM, Ramay SM, Saleem M. $BiFeO_3$ and La doped $BiFeO_3$ nano-particles decorated anodic $Al_2O_3$ porous template fabricated with two step anodization. Mater Lett 2019;244:115–8. doi:10.1016/j.matlet.2019.02.061.

[17]  Akhlaq M, Manzoor S, Tayyaba S, Ashraf MW, Asif A. Fuzzy Analysis, Fabrication and Characterization of Nano-porous Anodic Aluminum Oxide Membrane for Bio-MEMS. Adv. Intell. Data Anal. Appl., Springer, Singapore; 2022, p. 341–53. doi:10.1007/978-981-16-5036-9_32.

[18]  Pervez MF, Mia MNH, Hossain S, Saha SMK, Ali MH, Sarker P, et al. Influence of total absorbed dose of gamma radiation on optical bandgap and structural properties of Mg-doped zinc oxide. Optik (Stuttg) 2018;162:140–50. doi:10.1016/j.ijleo.2018.02.063.

[19]  Mia MNH, Habiba U, Pervez MF, Kabir H, Nur S, Hossen MF, et al. Investigation of aluminum doping on structural and optical characteristics of sol–gel assisted spin-coated nano-structured zinc oxide thin films. Appl Phys A 2020;126:162. doi:10.1007/s00339-020-3332-z.

[20]  Mia MNH, Pervez MF, Hossain MK, Reefaz Rahman M, Uddin MJ, Al Mashud MA, et al. Influence of Mg content on tailoring optical bandgap of Mg-doped ZnO thin film prepared by sol-gel method. Results Phys 2017;7:2683–91. doi:10.1016/j.rinp.2017.07.047.

[21]  Larbot A, Alary JA, Guizard C, Cot L, Gillot J. New inorganic ultrafiltration membranes: Preparation and characterisation. Int J High Technol Ceram 1987;3:143–51. doi:10.1016/0267-3762(87)90034-8.

[22]  Saha S. Preparation of alumina by sol-gel process, its structures and properties. J Sol-Gel Sci Technol 1994;3:117–26. doi:10.1007/BF00486718.

[23]  Yoldas BE. Alumina gels that form porous transparent $Al_2O_3$. J Mater Sci 1975;10:1856–60. doi:10.1007/BF00754473.

[24]  Ksapabutr B, Gulari E, Wongkasemjit S. Sol–gel transition study and pyrolysis of alumina-based gels prepared from alumatrane precursor. Colloids Surfaces A Physicochem Eng Asp 2004;233:145–53. doi:10.1016/j.colsurfa.2003.11.019.

[25]  Leenaars AFM, Burggraaf AJ. The preparation and characterization of alumina membranes with ultra-fine pores. J Memb Sci 1985;24:245–60. doi:10.1016/S0376-7388(00)82243-7.

[26]  Ren C, Fang H, Gu J, Winnubst L, Chen C. Preparation and characterization of hydrophobic alumina planar membranes for water desalination. J Eur Ceram Soc 2015;35:723–30. doi:10.1016/j.jeurceramsoc.2014.07.012.

[27]  Liu S, Li K, Hughes R. Preparation of porous aluminium oxide ($Al_2O_3$) hollow fibre membranes by a combined phase-inversion and sintering method. Ceram Int 2003;29:875–81. doi:10.1016/S0272-8842(03)00030-0.

[28]  Zhu Z, Xiao J, He W, Wang T, Wei Z, Dong Y. A phase-inversion casting process for preparation of tubular porous alumina ceramic membranes. J Eur Ceram Soc 2015;35:3187–94. doi:10.1016/j.jeurceramsoc.2015.04.026.

[29]  Li L, Chen M, Dong Y, Dong X, Cerneaux S, Hampshire S, et al. A low-cost alumina-mullite composite hollow fiber ceramic membrane fabricated via phase-inversion and sintering method. J Eur Ceram Soc 2016;36:2057–66. doi:10.1016/j.jeurceramsoc.2016.02.020.

[30]  Coelho LL, Di Luccio M, Hotza D, de Fátima Peralta Muniz Moreira R, Moreira AC, Fernandes CP, et al. Tailoring asymmetric $Al_2O_3$ membranes by combining tape casting and phase inversion. J Memb Sci 2021;623:119056. doi:10.1016/j.memsci.2021.119056.

[31]   Wielage B, Alisch G, Lampke T, Nickel D. Anodizing – A Key for Surface Treatment of Aluminium. Key Eng Mater 2008;384:263–81. doi:10.4028/www.scientific.net/KEM.384.263.

[32]   Hossain MK, Khan MI, El-Denglawey A. A review on biomedical applications, prospects, and challenges of rare earth oxides. Appl Mater Today 2021;24:101104. doi:10.1016/j.apmt.2021.101104.

[33]   Hossain MK, Ahmed MH, Khan MI, Miah MS, Hossain S. Recent Progress of Rare Earth Oxides for Sensor, Detector, and Electronic Device Applications: A Review. ACS Appl Electron Mater 2021;3:4255–83. doi:10.1021/acsaelm.1c00703.

[34]   Guy Dunstan Bengough JMS. Improved process of protecting surfaces of aluminium of aluminium alloys., 1923.

[35]   Kape JM, Whittaker JA. Five-Year Exposure Tests on Anodized Aluminium in a Severe Industrial Atmosphere. Trans IMF 1965;43:106–11. doi:10.1080/00202967.1965.11869963.

[36]   Sacchi F. Recent Research on Anodizing and its Practical Implications. Trans IMF 1964;42:14–21. doi:10.1080/00202967.1964.11869906.

[37]   Tsyntsaru N. Porous anodized aluminium oxide: application outlooks. CHEMIJA 2016;27:17–23.

[38]   Sulka GD. Highly Ordered Anodic Porous Alumina Formation by Self-Organized Anodizing. Nanostructured Mater. Electrochem., Weinheim, Germany: Wiley-VCH Verlag GmbH & Co. KGaA; n.d., p. 1–116. doi:10.1002/9783527621507.ch1.

[39]   Zhou F. Growth mechanism of porous anodic films on aluminium. University of Manchester, 2011.

[40]   Poinern GEJ, Ali N, Fawcett D. Progress in Nano-Engineered Anodic Aluminum Oxide Membrane Development. Materials (Basel) 2011;4:487–526. doi:10.3390/ma4030487.

[41]   Lee W, Park S-J. Porous Anodic Aluminum Oxide: Anodization and Templated Synthesis of Functional Nanostructures. Chem Rev 2014;114:7487–556. doi:10.1021/cr500002z.

[42]   Wu Z, Richter C, Menon L. A Study of Anodization Process during Pore Formation in Nanoporous Alumina Templates. J Electrochem Soc 2007;154:E8. doi:10.1149/1.2382671.

[43]   Lee W, Ji R, Gösele U, Nielsch K. Fast fabrication of long-range ordered porous alumina membranes by hard anodization. Nat Mater 2006;5:741–7. doi:10.1038/nmat1717.

[44]   Shingubara S, Morimoto K, Sakaue H, Takahagi T. Self-Organization of a Porous Alumina Nanohole Array Using a Sulfuric/Oxalic Acid Mixture as Electrolyte. Electrochem Solid-State Lett 2004;7:E15. doi:10.1149/1.1644353.

[45]   Masuda H, Tanaka H, Baba N. Preparation of Porous Material by Replacing Microstructure of Anodic Alumina Film with Metal. Chem Lett 1990;19:621–2. doi:10.1246/cl.1990.621.

[46]   Masuda H, Satoh M. Fabrication of Gold Nanodot Array Using Anodic Porous Alumina as an Evaporation Mask. Jpn J Appl Phys 1996;35:L126–9. doi:10.1143/JJAP.35.L126.

[47]   Kushwaha MK. A comparative Study of Different Electrolytes for Obtaining Thick and Well-ordered nano-porous Anodic Aluminium Oxide (AAO) Films. Procedia Mater Sci 2014;5:1266–73. doi:10.1016/j.mspro.2014.07.438.

[48]   Tu GC, Huang LY. Hard anodizing of 2024 aluminium alloy using pulsed DC and AC power. Trans IMF 1987;65:60–6. doi:10.1080/00202967.1987.11870772.

[49]   Lee W, Schwirn K, Steinhart M, Pippel E, Scholz R, Gösele U. Structural engineering of nanoporous anodic aluminium oxide by pulse anodization of aluminium. Nat Nanotechnol 2008;3:234–9. doi:10.1038/nnano.2008.54.

[50]   Losic D, Lillo M, Losic D. Porous Alumina with Shaped Pore Geometries and Complex Pore Architectures Fabricated by Cyclic Anodization. Small 2009;5:1392–7. doi:10.1002/smll.200801645.

[51]   Iglesias-Rubianes L, Garcia-Vergara SJ, Skeldon P, Thompson GE, Ferguson J, Beneke M. Cyclic oxidation processes during anodizing of Al–Cu alloys. Electrochim Acta 2007;52:7148–57. doi:10.1016/j.electacta.2007.05.052.

[52]   O'sullivan J and GW. The morphology and mechanism of formation of porous anodic films on aluminium. Proc R Soc London A Math Phys Sci 1970;317:511–43. doi:10.1098/rspa.1970.0129.

[53]   BELWALKAR A, GRASING E, VANGEERTRUYDEN W, HUANG Z, MISIOLEK W. Effect of processing parameters on pore structure and thickness of anodic aluminum oxide (AAO) tubular membranes. J Memb Sci 2008;319:192–8. doi:10.1016/j.memsci.2008.03.044.

[54]   Sulka GD, Parkoła KG. Temperature influence on well-ordered nanopore structures grown by anodization of aluminium in sulphuric acid. Electrochim Acta 2007;52:1880–8. doi:10.1016/j.electacta.2006.07.053.

[55]   Schwirn K, Lee W, Hillebrand R, Steinhart M, Nielsch K, Gösele U. Self-Ordered Anodic Aluminum Oxide Formed by H 2 SO 4 Hard Anodization. ACS Nano 2008;2:302–10. doi:10.1021/nn7001322.

[56]   Furneaux RC, Rigby WR, Davidson AP. The formation of controlled-porosity membranes from anodically oxidized aluminium. Nature 1989;337:147–9. doi:10.1038/337147a0.

[57]   Zaraska L, Sulka GD, Jaskuła M. The effect of n-alcohols on porous anodic alumina formed by self-organized two-step anodizing of aluminum in phosphoric acid. Surf Coatings Technol 2010;204:1729–37. doi:10.1016/j.surfcoat.2009.10.051.

[58]   Buijnsters JG, Zhong R, Tsyntsaru N, Celis J-P. Surface Wettability of Macroporous Anodized Aluminum Oxide.

ACS Appl Mater Interfaces 2013;5:3224–33. doi:10.1021/am4001425.

[59] Lee W, Kim J-C, Gösele U. Spontaneous Current Oscillations during Hard Anodization of Aluminum under Potentiostatic Conditions. Adv Funct Mater 2010;20:21–7. doi:10.1002/adfm.200901213.

[60] Rahman M, Garcia-Caurel E, Santos A, Marsal LF, Pallarès J, Ferré-Borrull J. Effect of the anodization voltage on the pore-widening rate of nanoporous anodic alumina. Nanoscale Res Lett 2012;7:474. doi:10.1186/1556-276X-7-474.

[61] Ferré-Borrull J, Rahman MM, Pallarès J, Marsal LF. Tuning nanoporous anodic alumina distributed-Bragg reflectors with the number of anodization cycles and the anodization temperature. Nanoscale Res Lett 2014;9:416. doi:10.1186/1556-276X-9-416.

[62] Chung CK, Zhou RX, Liu TY, Chang WT. Hybrid pulse anodization for the fabrication of porous anodic alumina films from commercial purity (99%) aluminum at room temperature. Nanotechnology 2009;20:055301. doi:10.1088/0957-4484/20/5/055301.

[63] Raspal V, Awitor KO, Massard C, Feschet-Chassot E, Bokalawela RSP, Johnson MB. Nanoporous Surface Wetting Behavior: The Line Tension Influence. Langmuir 2012;28:11064–71. doi:10.1021/la301201k.

[64] Sulka GD, Parkoła KG. Anodising potential influence on well-ordered nanostructures formed by anodisation of aluminium in sulphuric acid. Thin Solid Films 2006;515:338–45. doi:10.1016/j.tsf.2005.12.094.

[65] Sulka GD, Hnida K. Distributed Bragg reflector based on porous anodic alumina fabricated by pulse anodization. Nanotechnology 2012;23:075303. doi:10.1088/0957-4484/23/7/075303.

[66] Kozhukhova AE, du Preez SP, Bessarabov DG. Preparation of anodized aluminium oxide at high temperatures using low purity aluminium (Al6082). Surf Coatings Technol 2019;378:124970. doi:10.1016/j.surfcoat.2019.124970.

[67] Erdogan P, Yuksel B, Birol Y. Effect of chemical etching on the morphology of anodic aluminum oxides in the two-step anodization process. Appl Surf Sci 2012;258:4544–50. doi:10.1016/j.apsusc.2012.01.025.

[68] Ateş S, Baran E, Yazıcı B. The nanoporous anodic alumina oxide formed by two-step anodization. Thin Solid Films 2018;648:94–102. doi:10.1016/j.tsf.2018.01.013.

[69] Kovaleva EG, Molochnikov LS, Tambasova D, Marek A, Chestnut M, Osipova VA, et al. Electrostatic properties of inner nanopore surfaces of anodic aluminum oxide membranes upon high temperature annealing revealed by EPR of pH-sensitive spin probes and labels. J Memb Sci 2020;604:118084. doi:10.1016/j.memsci.2020.118084.

[70] Yang J, Wang J, Wang C-W, He X, Li Y, Chen J-B, et al. Intermediate wetting states on nanoporous structures of anodic aluminum oxide surfaces. Thin Solid Films 2014;562:353–60. doi:10.1016/j.tsf.2014.04.020.

[71] Zhang H, Yin L, Shi S, Liu X, Wang Y, Wang F. Facile and fast fabrication method for mechanically robust superhydrophobic surface on aluminum foil. Microelectron Eng 2015;141:238–42. doi:10.1016/j.mee.2015.03.048.

[72] Ye J, Yin Q, Zhou Y. Superhydrophilicity of anodic aluminum oxide films: From "honeycomb" to "bird's nest." Thin Solid Films 2009;517:6012–5. doi:10.1016/j.tsf.2009.04.042.

[73] Zhang W, Huang L, Zi C, Cai Y, Zhang Y, Zhou X, et al. Wettability of porous anodic aluminium oxide membranes with three-dimensional, layered nanostructures. J Porous Mater 2018;25:1707–14. doi:10.1007/s10934-018-0584-5.

[74] Ko S, Lee D, Jee S, Park H, Lee K, Hwang W. Mechanical properties and residual stress in porous anodic alumina structures. Thin Solid Films 2006;515:1932–7. doi:10.1016/j.tsf.2006.07.169.

[75] Sundararajan M, Devarajan M, Jaafar M. Investigation of surface and mechanical properties of Anodic Aluminium Oxide (AAO) developed on Al substrate for an electronic package enclosure. Surf Coatings Technol 2020;401:126273. doi:10.1016/j.surfcoat.2020.126273.

[76] Osmanbeyoglu HU, Hur TB, Kim HK. Thin alumina nanoporous membranes for similar size biomolecule separation. J Memb Sci 2009;343:1–6. doi:10.1016/j.memsci.2009.07.027.

[77] Brüggemann D. Nanoporous Aluminium Oxide Membranes as Cell Interfaces. J Nanomater 2013;2013:1–18. doi:10.1155/2013/460870.

[78] Poinern, G.E.J., Le, X.T., Hager, M., Becker, T. and Fawcett D. Electrochemical synthesis, characterisation, and preliminary biological evaluation of an anodic aluminium oxide membrane with a pore size of 100 nanometres for a potential cell culture substrate. Am J Biomed Eng 2013;3:119–31. doi:10.5923/j.ajbe.20130306.01.

[79] Mao Y, Park T-J, Zhang F, Zhou H, Wong SS. Environmentally Friendly Methodologies of Nanostructure Synthesis. Small 2007;3:1122–39. doi:10.1002/smll.200700048.

[80] Shaban M, Mustafa M, Khan AAP. Hexagonal diameter in cadmium sulfide/anodic alumina nanoporous bi-layer membrane by a sol–gel spin coating and their sensing application. Appl Phys A Mater Sci Process 2020;126. doi:10.1007/s00339-020-3371-5.

[81] Ruiz-Clavijo A, Caballero-Calero O, Martín-González M. Revisiting anodic alumina templates: From fabrication to applications. Nanoscale 2021;13:2227–65. doi:10.1039/d0nr07582e.

[82] Li W, Zhang J, Shen T, Jones GA, Grundy PJ. Magnetic nanowires fabricated by anodic aluminum oxide template- a brief review. Sci China Physics, Mech Astron 2011;54:1181–9. doi:10.1007/s11433-011-4371-4.

[83]   Tomassi P, Buczko Z. Aluminum Anodic Oxide AAO as a Template for Formation of Metal Nanostructures. Electroplat. Nanostructures, InTech; 2015. doi:10.5772/61263.

[84]   Yi G, Schwarzacher W. Single crystal superconductor nanowires by electrodeposition. Appl Phys Lett 1999;74:1746–8. doi:10.1063/1.123675.

[85]   Zhang Y, Li G, Wu Y, Zhang B, Song W, Zhang L. Antimony Nanowire Arrays Fabricated by Pulsed Electrodeposition in Anodic Alumina Membranes. Adv Mater 2002;14:1227–30. doi:10.1002/1521-4095(20020903)14:17<1227::AID-ADMA1227>3.0.CO;2-2.

[86]   Nielsch K, Müller F, Li A-P, Gösele U. Uniform Nickel Deposition into Ordered Alumina Pores by Pulsed Electrodeposition. Adv Mater 2000;12:582–6. doi:10.1002/(SICI)1521-4095(200004)12:8<582::AID-ADMA582>3.0.CO;2-3.

[87]   Pan H, Sun H, Poh C, Feng Y, Lin J. Single-crystal growth of metallic nanowires with preferred orientation. Nanotechnology 2005;16:1559–64. doi:10.1088/0957-4484/16/9/025.

[88]   Thongmee S, Pang HL, Ding J, Yi JB, Lin JY. Fabrication and magnetic properties of metal nanowires via AAO templates. 2008 2nd IEEE Int. Nanoelectron. Conf., IEEE; 2008, p. 1116–20. doi:10.1109/INEC.2008.4585678.

[89]   Yang R, Sui C, Gong J, Qu L. Silver nanowires prepared by modified AAO template method. Mater Lett 2007;61:900–3. doi:10.1016/j.matlet.2006.06.009.

[90]   Wu Z, Zhang Y, Du K. A simple and efficient combined AC–DC electrodeposition method for fabrication of highly ordered Au nanowires in AAO template. Appl Surf Sci 2013;265:149–56. doi:10.1016/j.apsusc.2012.10.154.

[91]   Ganapathi A, Swaminathan P, Neelakantan L. Anodic Aluminum Oxide Template Assisted Synthesis of Copper Nanowires using a Galvanic Displacement Process for Electrochemical Denitrification. ACS Appl Nano Mater 2019;2:5981–8. doi:10.1021/acsanm.9b01409.

[92]   Zhang C, Lu Y, Zhao B, Hao Y, Liu Y. Facile fabrication of Ag dendrite-integrated anodic aluminum oxide membrane as effective three-dimensional SERS substrate. Appl Surf Sci 2016;377:167–73. doi:10.1016/j.apsusc.2016.03.132.

[93]   Lai C-H, Chang C-W, Tseng T-Y. Size-dependent field-emission characteristics of ZnO nanowires grown by porous anodic aluminum oxide templates assistance. Thin Solid Films 2010;518:7283–6. doi:10.1016/j.tsf.2010.04.091.

[94]   Li H, Zhang H, Wu H, Hao J, Liu C. Tree-like structures of InN nanoparticles on agminated anodic aluminum oxide by plasma-assisted reactive evaporation. Appl Surf Sci 2020;503. doi:10.1016/j.apsusc.2019.144309.

[95]   Maurya MR, Toutam V. Fast response UV detection based on waveguide characteristics of vertically grown ZnO nanorods partially embedded in anodic alumina template. Nanotechnology 2019;30. doi:10.1088/1361-6528/aaf545.

[96]   Liu C-J, Chen S-Y, Shih L-J, Huang H-J. Fabrication of nanotubules of thermoelectric γ-Na0.7CoO2 using porous aluminum oxide membrane as supporting template. Mater Chem Phys 2010;119:424–7. doi:10.1016/j.matchemphys.2009.09.016.

[97]   Tzaneva BR, Naydenov AI, Todorova SZ, Videkov VH, Milusheva VS, Stefanov PK. Cobalt electrodeposition in nanoporous anodic aluminium oxide for application as catalyst for methane combustion. Electrochim Acta 2016;191:192–9. doi:10.1016/j.electacta.2016.01.063.

[98]   Maghsodi A, Adlnasab L, Shabanian M, Javanbakht M. Optimization of effective parameters in the synthesis of nanopore anodic aluminum oxide membrane and arsenic removal by prepared magnetic iron oxide nanoparicles in anodic aluminum oxide membrane via ultrasonic-hydrothermal method. Ultrason Sonochem 2018;48:441–52. doi:10.1016/j.ultsonch.2018.07.003.

[99]   Nehra M, Dilbaghi AN, Singh V, Singhal NK, Kumar S. Highly Ordered and Crystalline Cu Nanowires in Anodic Aluminum Oxide Membranes for Biomedical Applications. Phys Status Solidi 2020;217:1900842. doi:10.1002/pssa.201900842.

[100]  Ghezghapan, S. M. S., & Saba M. Synthesizing high aspect ratio Aluminum Oxide nanowires from highly-ordered anodic self-assembled templates. 2018. doi:10.1149/osf.io/zejvp.

[101]  Esfandi F, Saramad S, Shahmirzadi MR. Characterizing and simulation the scintillation properties of zinc oxide nanowires in AAO membrane for medical imaging applications. J Instrum 2017;12:P07004–P07004. doi:10.1088/1748-0221/12/07/P07004.

[102]  Attiq-ur-rehman, Ashraf MW, Mahmood A, Rehman AU, Ramay SM, Saleem M. Growth of Zr–BiFeO3 nanostructures on two step anodized porous alumina for estimation of optical and dielectric response. Phys E Low-Dimensional Syst Nanostructures 2021;127. doi:10.1016/j.physe.2020.114513.

[103]  Lednický T, Bonyár A. Large Scale Fabrication of Ordered Gold Nanoparticle–Epoxy Surface Nanocomposites and Their Application as Label-Free Plasmonic DNA Biosensors. ACS Appl Mater Interfaces 2020;12:4804–14. doi:10.1021/acsami.9b20907.

[104]  Pourjafari D, Serrano T, Kharissov B, Peña Y, Gómez I. Template assisted synthesis of poly(3-hexylthiophene) nanorods and nanotubes: Growth mechanism and corresponding band gap. Int J Mater Res 2015;106:414–20. doi:10.3139/146.111196.

[105]   Gao X, Liu L, Birajdar B, Ziese M, Lee W, Alexe M, et al. High-density periodically ordered magnetic cobalt ferrite nanodot arrays by template-assisted pulsed laser deposition. Adv Funct Mater 2009;19:3450–5. doi:10.1002/adfm.200900422.

[106]   Chen P-L, Chang J-K, Kuo C-T, Pan F-M. Anodic aluminum oxide template assisted growth of vertically aligned carbon nanotube arrays by ECR-CVD 2004;13:1949–53. doi:10.1016/j.diamond.2004.05.007.

[107]   Xu T-N, Wu H-Z, Lao Y-F, Qiu D-J, Chen N-B, Dai N. Anodic-aluminium-oxide template-assisted growth of ZnO nanodots on Si (100) at low temperature. Chinese Phys Lett 2004;21:1327–9. doi:10.1088/0256-307X/21/7/040.

[108]   Yang W, Oh Y, Kim J, Kim H, Shin H, Moon J. Photoelectrochemical Properties of Vertically Aligned CuInS2 Nanorod Arrays Prepared via Template-Assisted Growth and Transfer. ACS Appl Mater Interfaces 2016;8:425–31. doi:10.1021/acsami.5b09241.

[109]   Su CC, Teng YC, Chang SH. Synthesize uniform carbon nanocoils by anodic aluminum oxide, 2010, p. 950–3. doi:10.1109/NEMS.2010.5592107.

[110]   W Shi, Y Shen, D Ge, M Xue, H Cao SH. Functionalized anodic aluminum oxide (AAO) membranes for affinity protein separation. J Memb Sci 2008;325:801–8. doi:10.1016/j.memsci.2008.09.003.

[111]   Lee, Sang Bok, David T. Mitchell, Lacramioara Trofin, Tarja K. Nevanen, Hans Söderlund and CRM. Antibody-based bio-nanotube membranes for enantiomeric drug separations. Science (80- ) 2002;296:2198–200. doi:10.1126/science.1071396.

[112]   Losic, Dusan, Martin A. Cole, Björn Dollmann, Krasimir Vasilev and HJG. Surface modification of nanoporous alumina membranes by plasma polymerization. Nanotechnology 2008;19:24570. doi:10.1088/0957-4484/19/24/245704.

[113]   Yamaguchi, Akira, Fumiaki Uejo, Takashi Yoda, Tatsuya Uchida, Yoshihiko Tanamura, Tomohisa Yamashita and NT. Self-assembly of a silica–surfactant nanocomposite in a porous alumina membrane. Nat Mater 2004;3:337–41. doi:10.1038/nmat1107.

[114]   Muhamad Waseem Ashraf, Muhammad Zahid Qureshi, Fozia Ghaffar, Shahzadi Tayyaba NA. Structural study of AAO membrane during the dialysis process. 2016 Int. Conf. Intell. Syst. Eng., 2016. doi:10.1109/INTELSE.2016.7475125.

[115]   Thormann A, Berthold L, Gšring P, Lelonek M, Heilmann A. Nanoporous Aluminum Oxide Membranes for Separation and Biofunctionalization. Procedia Eng 2012;44:1107–11. doi:10.1016/j.proeng.2012.08.693.

[116]   Khan Kasi A, Khan Kasi J, Afzulpurkar N, Bohez E, Tuantranont A, Mahaisavariya B. Novel anodic aluminum oxide (AAO) nanoporous membrane for wearable hemodialysis device. Int. Conf. Commun. Electron. 2010, IEEE; 2010, p. 98–101. doi:10.1109/ICCE.2010.5670689.

[117]   Yaw-JenChang W-T-C. Isolation and detection of exosomes via AAO membrane and QCM measurement. Microelectron Eng 2019;216. doi:10.1016/j.mee.2019.111094.

[118]   Liu Y, Lou J, Ni M, Song C, Wu J, Dasgupta NP, et al. Bioinspired Bifunctional Membrane for Efficient Clean Water Generation. ACS Appl Mater Interfaces 2016;8:772–9. doi:10.1021/acsami.5b09996.

[119]   Nikam SB, SK A. Enantioselective Separation Using Chiral Amino Acid Functionalized Polyfluorene Coated on Mesoporous Anodic Aluminum Oxide Membranes. Anal Chem 2020;92:6850–7. doi:10.1021/acs.analchem.9b04699.

[120]   Kim, Young Jo, John E. Jones, Hao Li, Helen Yampara-Iquise, Guolu Zheng, Charles A. Carson, Michael Cooperstock, Michael Sherman and QY. Three-dimensional (3-D) microfluidic-channel-based DNA biosensor for ultra-sensitive electrochemical detection. J Electroanal Chem 2013;702:72–8. doi:10.1016/j.jelechem.2013.04.021.

[121]   Song J, Oh H, Kong H, Jang J. Polyrhodanine modified anodic aluminum oxide membrane for heavy metal ions removal. J Hazard Mater 2011;187:311–7. doi:10.1016/j.jhazmat.2011.01.026.

[122]   Kasi, Ajab Khan, Jafar Khan Kasi and MB. Fabrication of mechanically stable AAO membrane with improved fluid permeation properties. Microelectron Eng 2018;187:95-100. doi:10.1016/j.mee.2017.11.019.

[123]   Chang, H-C., Y-H. Chen, A-T. Lo, S-S. Hung, S-L. Lin, I-N. Chang and J-CL. Modified nanoporous membranes on centrifugal microfluidic platforms for detecting heavy metal ions. Mater Res Innov 2014;18:S2-685. doi:10.1179/1432891714Z.000000000532.

[124]   Kim, Yunho, Misun Cha, Yosep Choi, Hyunsang Joo and JL. Electrokinetic separation of biomolecules through multiple nano-pores on membrane. Chem Phys Lett 2013;561:63–7. doi:10.1016/j.cplett.2013.01.018.

[125]   Phuong, NguyenThi, Anugrah Andisetiawan, Jeong Hwan Kim, Doo-Sun Choi, Kyung-Hyun Whang, Jeasun Nham, Yun Jung Lee, Yeong-Eun Yoo and JSY. Nano sand filter with functionalized nanoparticles embedded in anodic aluminum oxide templates. Sci Rep 2016;6:1–8.

[126]   Hun CW, Chiu Y-J, Luo Z, Chen CC, Chen SH. A new technique for batch production of tubular anodic aluminum oxide films for filtering applications. Appl Sci 2018;8. doi:10.3390/app8071055.

[127]   Hasan, Mahadi; Kasi, Ajab Khan; Kasi, Jafar Khan; Afzulpurkar N. Anodic Aluminum Oxide (AAO) to AAO Bonding and Their Application for Fabrication of 3D Microchannel. Nanosci Nanotechnol Lett 2012;4:569-573(5). doi:10.1166/nnl.2012.1354.

[128] Patel, Yatinkumar, Giedrius Janusas, Arvydas Palevicius and AV. Development of Nanoporous AAO Membrane for Nano Filtration Using the Acoustophoresis Method. Sensors 2020;20. doi:10.3390/s20143833.

[129] Ma Y, Kaczynski J, Ranacher C, Roshanghias A, Zauner M, Abasahl B. Nano-porous aluminum oxide membrane as filtration interface for optical gas sensor packaging. Microelectron Eng 2018;198:29–34. doi:10.1016/j.mee.2018.06.013.

[130] Ajab Khan Kasi, Jafar Khan Kasi, Mahadi Hasan, Nitin Afzulpurkar, Sirapat Pratontep, Supanit Pornitheeraphat AP. Fabrication of Low Cost Anodic Aluminum Oxide (AAO) Tubular Membrane and their Application for Hemodialysis. Adv Mater Res 2012;550–553. doi:10.4028/www.scientific.net/AMR.550-553.2040.

[131] Fang-Yu Wen, Po-Sheng Chen, Ting-Wei Liao Y-JJ. Microwell-assisted filtration with anodic aluminum oxide membrane for Raman analysis of algal cells. Algal Res 2018;33:412–8. doi:10.1016/j.algal.2018.06.022.

[132] Park, Yong, Sueon Kim, In Hyuk Jang, Young Suk Nam, Hiki Hong, Dukhyun Choi and WGL. Role of the electric field in selective ion filtration in nanostructures. Analyst 2016;141:1294–300. doi:10.1039/C5AN01980J.

[133] Santos A, Kumeria T, Losic D. Nanoporous anodic aluminum oxide for chemical sensing and biosensors. TrAC - Trends Anal Chem 2013;44:25–38. doi:10.1016/j.trac.2012.11.007.

[134] Gorokh G, Mozalev A, Solovei D, Khatko V, Llobet E, Correig X. Anodic formation of low-aspect-ratio porous alumina films for metal-oxide sensor application. Electrochim Acta 2006;52:1771–80. doi:10.1016/j.electacta.2006.01.081.

[135] Han N, Deng P, Chen J, Chai L, Gao H, Chen Y. Electrophoretic deposition of metal oxide films aimed for gas sensors application: The role of anodic aluminum oxide (AAO)/Al composite structure. Sensors Actuators, B Chem 2010;144:267–73. doi:10.1016/j.snb.2009.10.068.

[136] Lo P-H, Hong C, Lo S-C, Fang W. Implementation of inductive proximity sensor using nanoporous anodic aluminum oxide layer, 2011, p. 1871–4. doi:10.1109/TRANSDUCERS.2011.5969829.

[137] Hong C, Chu L, Lai W, Chiang A-S, Fang W. Implementation of a new capacitive touch sensor using the nanoporous anodic aluminum oxide (np-AAO) structure. IEEE Sens J 2011;11:3409–16. doi:10.1109/JSEN.2011.2160255.

[138] Balde M, Vena A, Sorli B. Fabrication of porous anodic aluminium oxide layers on paper for humidity sensors. Sensors Actuators, B Chem 2015;220:829–39. doi:10.1016/j.snb.2015.05.053.

[139] Jeong SH, Im HL, Hong S, Park H, Baek J, Park DH, et al. Massive, eco-friendly, and facile fabrication of multi-functional anodic aluminum oxides: application to nanoporous templates and sensing platforms. RSC Adv 2017;7:4518–30. doi:10.1039/C6RA25201J.

[140] Kumeria T, Kurkuri M, Diener K, Zhang C, Parkinson L, Losic D. Reflectometric interference biosensing using nanopores: Integration into microfluidics. vol. 8204, 2011. doi:10.1117/12.903217.

[141] Chen W, Gui X, Liang B, Yang R, Zheng Y, Zhao C, et al. Structural Engineering for High Sensitivity, Ultrathin Pressure Sensors Based on Wrinkled Graphene and Anodic Aluminum Oxide Membrane. ACS Appl Mater Interfaces 2017;9:24111–7. doi:10.1021/acsami.7b05515.

[142] Peng D, Chen J, Jiao L, Liu Y. A fast-responding semi-transparent pressure-sensitive paint based on through-hole anodized aluminum oxide membrane. Sensors Actuators A Phys 2018;274:10–8. doi:10.1016/j.sna.2018.02.026.

[143] Pinkhasova P, Chen H, Verhoeven MWGM (Tiny), Sukhishvili S, Du H. Thermally annealed Ag nanoparticles on anodized aluminium oxide for SERS sensing. RSC Adv 2013;3:17954. doi:10.1039/c3ra43808b.

[144] He Y, Li X, Que L. A transparent nanostructured optical biosensor. J Biomed Nanotechnol 2014;10:767–74. doi:10.1166/jbn.2014.1769.

[145] Huang C-H, Lin H-Y, Chen S, Liu C-Y, Chui H-C, Tzeng Y. Electrochemically fabricated self-aligned 2-D silver/alumina arrays as reliable SERS sensors. Opt Express 2011;19:11441–50. doi:10.1364/OE.19.011441.

[146] Umh H-N, Shin HH, Yi J, Kim Y. Fabrication of gold nanowires (GNW) using aluminum anodic oxide (AAO) as a metal-ion sensor. Korean J Chem Eng 2015;32:299–302. doi:10.1007/s11814-014-0201-5.

[147] Song W, Gan B, Jiang T, Zhang Y, Yu A, Yuan H, et al. Nanopillar Arrayed Triboelectric Nanogenerator as a Self-Powered Sensitive Sensor for a Sleep Monitoring System. ACS Nano 2016;10:8097–103. doi:10.1021/acsnano.6b04344.

[148] Mondal S, Kim SJ, Choi C-G. Honeycomb-like MoS2 Nanotube Array-Based Wearable Sensors for Noninvasive Detection of Human Skin Moisture. ACS Appl Mater Interfaces 2020;12:17029–38. doi:10.1021/acsami.9b22915.

[149] Wang G, Wang J, Li S-Y, Zhang J-W, Wang C-W. One-dimensional alumina photonic crystals with a narrow band gap and their applications to high-sensitivity concentration sensor and photoluminescence enhancement. Superlattices Microstruct 2015;86:546–51. doi:10.1016/j.spmi.2015.08.004.

[150] Fan Y, Ding Y, Zhang Y, Ma H, He Y, Sun S. A SiO2-coated nanoporous alumina membrane for stable label-free waveguide biosensing. RSC Adv 2014;4:62987–95. doi:10.1039/c4ra08839e.

[151] Podgolin SK, Petukhov DI, Dorofeev SG, Eliseev AA. Anodic alumina membrane capacitive sensors for detection of vapors. Talanta 2020;219:121248. doi:10.1016/j.talanta.2020.121248.

[152] Anik MI, Hossain MK, Hossain I, Ahmed I, Doha RM. Biomedical applications of magnetic nanoparticles. In:

Ehrmann A, Nguyen TA, Ahmadi M, Farmani A, Nguyen-Tri P, editors. Magn. Nanoparticle-Based Hybrid Mater. 1st ed., Sawston, United Kingdom: Elsevier; 2021, p. 463–97. doi:10.1016/B978-0-12-823688-8.00002-8.

[153] Anik MI, Hossain MK, Hossain I, Mahfuz AMUB, Rahman MT, Ahmed I. Recent progress of magnetic nanoparticles in biomedical applications: A review. Nano Sel 2021;2:1146–86. doi:10.1002/nano.202000162.

[154] Gupta Manish SV. Targeted drug delivery system: A Review. Res J Chem Sci 2011;2:135–8.

[155] Sun X, Jiang L, Wang C, Sun S, Mei L, Huang L. Systematic investigation of intracellular trafficking behavior of one-dimensional alumina nanotubes. J Mater Chem B 2019;7:2043–53. doi:10.1039/c8tb03349h.

[156] Li L, Chen D, Zhang Y, Deng Z, Ren X, Meng X, et al. Magnetic and fluorescent multifunctional chitosan nanoparticles as a smart drug delivery system. Nanotechnology 2007;18:405102. doi:10.1088/0957-4484/18/40/405102.

[157] Hou P, Liu C, Shi C, Cheng H. Carbon nanotubes prepared by anodic aluminum oxide template method. Chinese Sci Bull 2012;57:187–204. doi:10.1007/s11434-011-4892-2.

[158] Losic D, Simovic S. Self-ordered nanopore and nanotube platforms for drug delivery applications. Expert Opin Drug Deliv 2009;6:1363–81. doi:10.1517/17425240903300857.

[159] Sinn Aw M, Kurian M, Losic D. Non-eroding drug-releasing implants with ordered nanoporous and nanotubular structures: Concepts for controlling drug release. Biomater Sci 2014;2:10–34. doi:10.1039/c3bm60196j.

[160] Yin J, Cui Y, Yang G, Wang H. Molecularly imprinted nanotubes for enantioselective drug delivery and controlled release. Chem Commun 2010;46:7688. doi:10.1039/c0cc01782e.

[161] Mushtaq F, Torlakcik H, Hoop M, Jang B, Carlson F, Grunow T, et al. Motile Piezoelectric Nanoeels for Targeted Drug Delivery. Adv Funct Mater 2019;29:1808135. doi:10.1002/adfm.201808135.

[162] Aw MS, Simovic S, Addai-Mensah J, Losic D. Polymeric micelles in porous and nanotubular implants as a new system for extended delivery of poorly soluble drugs. J Mater Chem 2011;21:7082–9. doi:10.1039/c0jm04307a.

[163] La Flamme KE, Latempa TJ, Grimes CA, Desai TA. The Effects of Cell Density and Device Arrangement on the Behavior of Macroencapsulated β-Cells. Cell Transplant 2007;16:765–74. doi:10.3727/000000007783465262.

[164] Kang H-J, Kim DJ, Park S-J, Yoo J-B, Ryu YS. Controlled drug release using nanoporous anodic aluminum oxide on stent. Thin Solid Films 2007;515:5184–7. doi:10.1016/j.tsf.2006.10.029.

[165] Simovic S, Losic D, Vasilev K. Controlled drug release from porous materials by plasma polymer deposition. Chem Commun 2010;46:1317. doi:10.1039/b919840g.

[166] Kwak D-H, Yoo J-B, Kim DJ. Drug Release behavior from nanoporous anodic aluminum oxide. J Nanosci Nanotechnol 2010;10:345–8. doi:10.1166/jnn.2010.1531.

[167] Park SB, Joo Y-H, Kim H, Ryu W, Park Y-I. Biodegradation-tunable mesoporous silica nanorods for controlled drug delivery. Mater Sci Eng C 2015;50:64–73. doi:10.1016/j.msec.2015.01.073.

[168] Niamlaem M, Phuakkong O, Garrigue P, Goudeau B, Ravaine V, Kuhn A, et al. Asymmetric Modification of Carbon Nanotube Arrays with Thermoresponsive Hydrogel for Controlled Delivery. ACS Appl Mater Interfaces 2020;12:23378–87. doi:10.1021/acsami.0c01017.

[169] Law CS, Santos A, Kumeria T, Losic D. Engineered therapeutic-releasing nanoporous anodic alumina-aluminum wires with extended release of therapeutics. ACS Appl Mater Interfaces 2015;3:3846–53. doi:10.1021/am5091963.

[170] Noh K, Brammer KS, Choi C, Kim SH, Frandsen CJ, Jin S. A New Nano-Platform for Drug Release via Nanotubular Aluminum Oxide. J Biomater Nanobiotechnol 2011;02:226–33. doi:10.4236/jbnb.2011.23028.

[171] Song C, Ben-Shlomo G, Que L. A Multifunctional Smart Soft Contact Lens Device Enabled by Nanopore Thin Film for Glaucoma Diagnostics and In Situ Drug Delivery. J Microelectromechanical Syst 2019;28:810–6. doi:10.1109/JMEMS.2019.2927232.

[172] Saji VS, Kumeria T, Gulati K, Prideaux M, Rahman S, Alsawat M, et al. Localized drug delivery of selenium (Se) using nanoporous anodic aluminium oxide for bone implants. J Mater Chem B 2015;3:7090–8. doi:10.1039/c5tb00125k.

[173] Kim S, Ozalp EI, Darwish M, Weldon JA. Electrically gated nanoporous membranes for smart molecular flow control. Nanoscale 2018;10:20740–7. doi:10.1039/c8nr05906c.

[174] Aw MS, Simovic S, Dhiraj K, Addai-Mensah J, Losic D. The loading and release property of nanoporous anodic alumina for delivery of drugs and drug carriers, 2010, p. 143–5. doi:10.1109/ICONN.2010.6045213.

[175] Thorat SB, Diaspro A, Salerno M. In vitro investigation of coupling-agent-free dental restorative composite based on nano-porous alumina fillers. J Dent 2014;42:279–86. doi:10.1016/j.jdent.2013.12.001.

[176] Fazli-Abukheyli R, Rahimi MR, Ghaedi M. Electrospinning coating of nanoporous anodic alumina for controlling the drug release: Drug release study and modeling. J Drug Deliv Sci Technol 2019;54. doi:10.1016/j.jddst.2019.101247.

[177] Jeon G, Yang SY, Byun J, Kim JK. Electrically actuatable smart nanoporous membrane for pulsatile drug release. Nano Lett 2011;11:1284–8. doi:10.1021/nl104329y.

[178] Chen G, Chen R, Zou C, Yang D, Chen Z-S. Fragmented polymer nanotubes from sonication-induced scission with

a thermo-responsive gating system for anti-cancer drug delivery. J Mater Chem B 2014;2:1327–34. doi:10.1039/C3TB21512A.

[179] Buyukserin F, Altuntas S, Aslim B. Fabrication and modification of composite silica nano test tubes for targeted drug delivery. RSC Adv 2014;4:23535–9. doi:10.1039/C4RA00871E.

[180] Hong C, Tang T-T, Hung C-Y, Pan R-P, Fang W. Liquid crystal alignment in nanoporous anodic aluminum oxide layer for LCD panel applications. Nanotechnology 2010;21:285201. doi:10.1088/0957-4484/21/28/285201.

[181] Zhou L, Tan Y, Ji D, Zhu B, Zhang P, Xu J, et al. Self-assembly of highly efficient, broadband plasmonic absorbers for solar steam generation. Sci Adv 2016;2:e1501227. doi:10.1126/sciadv.1501227.

[182] Davoodi E, Zhianmanesh M, Montazerian H, Milani AS, Hoorfar M. Nano-porous anodic alumina: fundamentals and applications in tissue engineering. J Mater Sci Mater Med 2020;31:60. doi:10.1007/s10856-020-06398-2.

[183] Park JS, Moon D, Kim J-S, Lee JS. Cell Adhesion and Growth on the Anodized Aluminum Oxide Membrane. J Biomed Nanotechnol 2016;12:575–80. doi:10.1166/jbn.2016.2192.

[184] Shi Y. Inhibiting the Growth of Lithium Dendrites by Employing the Anodic Aluminum Oxide Membrane. Northeastern University. ProQuest Dissertations Publishing,10928740., 2018.

[185] Mohan Raj R, Raj V. Fabrication of superhydrophobic coatings for combating bacterial colonization on Al with relevance to marine and medical applications. J Coatings Technol Res 2018;15:51–64. doi:10.1007/s11998-017-9945-2.

[186] Bunge F, van den Driesche S, Vellekoop MJ. PDMS-free microfluidic cell culture with integrated gas supply through a porous membrane of anodized aluminum oxide. Biomed Microdevices 2018;20:98. doi:10.1007/s10544-018-0343-z.

[187] Ingham CJ, ter Maat J, de Vos WM. Where bio meets nano: The many uses for nanoporous aluminum oxide in biotechnology. Biotechnol Adv 2012;30:1089–99. doi:10.1016/j.biotechadv.2011.08.005.

[188] Baskar, S., Vijayan, V., Saravanan, S., Balan, A. V., & Antony AG. Effect of $Al_2O_3$, aluminium alloy and fly ash for making engine component. Int J Mech Eng Technol 2018;9:91–96.

[189] Zhu J, Chen C, Lu Y, Zang J, Jiang M, Kim D, et al. Highly porous polyacrylonitrile/graphene oxide membrane separator exhibiting excellent anti-self-discharge feature for high-performance lithium–sulfur batteries. Carbon N Y 2016;101:272–80. doi:10.1016/j.carbon.2016.02.007.

[190] Lee Y, Kim HJ, Kim D-K. Power Generation from Concentration Gradient by Reverse Electrodialysis in Anisotropic Nanoporous Anodic Aluminum Oxide Membranes. Energies 2020;13:904. doi:10.3390/en13040904.

[191] Zarei H, Saramad S. The radiation gas detectors with novel nanoporous converter for medical imaging applications. J Instrum 2018;13:C02053–C02053. doi:10.1088/1748-0221/13/02/C02053.

[192] Sarno M, Tamburrano A, Arurault L, Fontorbes S, Pantani R, Datas L, et al. Electrical conductivity of carbon nanotubes grown inside a mesoporous anodic aluminium oxide membrane. Carbon N Y 2013;55:10–22. doi:10.1016/j.carbon.2012.10.063.

[193] Gloukhovski R, Freger V, Tsur Y. A Novel Composite Nafion/Anodized Aluminium Oxide Proton Exchange Membrane. Fuel Cells 2016;16:434–43. doi:10.1002/fuce.201500166.

[194] Hashimoto H, Kojima S, Sasaki T, Asoh H. α-Alumina membrane having a hierarchical structure of straight macropores and mesopores inside the pore wall. J Eur Ceram Soc 2018;38:1836–40. doi:10.1016/j.jeurceramsoc.2017.11.032.

[195] Chen W, Wang X, He G, Li T, Gong X, Wu X. Anion exchange membrane with well-ordered arrays of ionic channels based on a porous anodic aluminium oxide template. J Appl Electrochem 2018;48:1151–61. doi:10.1007/s10800-018-1214-2.

[196] Wang S, Tian Y, Wang C, Hang C, Zhang H, Huang Y, et al. One-Step Fabrication of Copper Nanopillar Array-Filled AAO Films by Pulse Electrodeposition for Anisotropic Thermal Conductive Interconnectors. ACS Omega 2019;4:6092–6. doi:10.1021/acsomega.8b03533.

[197] Fu L, Wang Y, Jiang J, Lu B, Zhai J. Sandwich "Ion Pool"-Structured Power Gating for Salinity Gradient Generation Devices. ACS Appl Mater Interfaces 2021;13:35197–206. doi:10.1021/acsami.1c10183.

[198] Li H, Wu L, Zhang H, Dai W, Hao J, Wu H, et al. Self-Assembly of Carbon Black/AAO Templates on Nanoporous Si for Broadband Infrared Absorption. ACS Appl Mater Interfaces 2020;12:4081–7. doi:10.1021/acsami.9b19107.

[199] Li J, Wei H, Zhao K, Wang M, Chen D, Chen M. Effect of anodizing temperature and organic acid addition on the structure and corrosion resistance of anodic aluminum oxide films. Thin Solid Films 2020;713:138359. doi:10.1016/j.tsf.2020.138359.